\pdfoutput=1
\documentclass[preprint,aps,prd,showpacs,nofootinbib,superscriptaddress,tightenlines,12pt]{revtex4}
\usepackage[usenames]{color}
\usepackage{amsfonts}
\usepackage{mathrsfs}
\usepackage{graphicx}
\usepackage{epstopdf}
\usepackage{amsmath}
\usepackage{amssymb}
\usepackage{bm}
\usepackage{bbm}
\usepackage{multirow}
\usepackage{alltt}
\usepackage{fancyvrb}
\usepackage[T1]{fontenc}
\usepackage{alltt}
\usepackage{booktabs,tabularx}
\usepackage[stable]{footmisc}
\usepackage{hyperref}

\definecolor{Section}{rgb}{0,0,0}
\definecolor{Code}{rgb}{0.444444,0.222222,0.}

\def\subsec#1{\vspace{.3cm}{\color{Section}\noindent#1}\vspace{.3cm}}

\newcommand{\PreserveBackslash}[1]{\let\temp=\\#1\let\\=\temp}
\newcolumntype{C}[1]{>{\PreserveBackslash\centering}p{#1}}
\newcolumntype{R}[1]{>{\PreserveBackslash\raggedleft}p{#1}}
\newcolumntype{L}[1]{>{\PreserveBackslash\raggedright}p{#1}}

\begin{document}
\graphicspath{{eps/},{eps/Install/},{eps/FormLink/},{eps/FeynCalcFormLink/},{eps/Matteo/},{eps/RunForm/}}

\title{\textsc{FormLink}/\textsc{FeynCalcFormLink}

Embedding \textsc{FORM} in \textit{Mathematica} and \textsc{FeynCalc}}

\author{Feng Feng\footnote{E-mail: fengf@ihep.ac.cn} }
\affiliation{Center for High Energy Physics, Peking University, Beijing 100871, China\vspace{0.2cm}}
\author{Rolf Mertig\footnote{E-mail: rolf@mertig.com} }
\affiliation{GluonVision GmbH, B\"otzowstrasse 10, 10407 Berlin,
Germany\vspace{0.2cm}}

\date{\today}

\begin{abstract}
\textsc{FORM}, a symbolic manipulation system, has been widely used in a lot of
calculations for High Energy Physics due to its high performance and efficient design.
\textit{Mathematica}, another computational software program, has also widely
been used, but more for reasons of generality and user-friendliness than for speed.
Especially calculations involving tensors and noncommutative operations
like calculating Dirac traces can be rather slow in \textit{Mathematica}, compared to \textsc{FORM}.

In this article we describe \textsc{FormLink} and \textsc{FeynCalcFormLink}, two
\textit{Mathematica} packages to link \textit{Mathematica} and \textsc{FeynCalc} with \textsc{FORM}.
\textsc{FormLink} can be used without \textsc{FeynCalc} and
\textsc{FeynCalcFormLink}, which is an extension loading \textsc{FormLink} and \textsc{FeynCalc} automatically.

With these two packages the impressive speed and other special features of \textsc{FORM}
get embedded into the generality of \textit{Mathematica} and \textsc{FeynCalc} in a simple manner.

\textsc{FeynCalcFormLink} provides a \textsc{FORM}-based turbo for {\sc
FeynCalc}, making it much more efficient.
\textsc{FormLink} turns \textit{Mathematica} into an editor and code organizer for \textsc{FORM}.
\end{abstract}

\pacs{\it  12.38.Bx}

\maketitle
{\bf PROGRAM SUMMARY}
\begin{itemize}

\item[]{\em Title of program:} {\tt \textsc{FormLink/FeynCalcFormLink}}
\item[]{\em Available from:}  {\tt \url{http://www.feyncalc.org/formlink/}}
\item[]{\em Programming language:} {\tt \textit{Mathematica} 7, 8 or 9; \textsc{C}; \textsc{FORM} 4}
\item[]{\em Computer:} Any computer running \textit{Mathematica} and \textsc{FORM}.
\item[]{\em Operating system:} Linux, Windows, Mac OS X.
\item[]{\em External routines:} \textsc{FORM}, \textsc{C}, \textsc{FeynCalc}.
\item[]{\em Keywords:} \textit{Mathematica}, \textit{MathLink}, \textsc{FORM}, FeynCalc, FormLink, FeynCalcFormLink.
\item[]{\em Nature of physical problem:}
The functionality of {\sc FORM} is restricted compared with \textit{Mathematica}, which
is broader, while its speed is slower for larger calculations involving tensors and noncommutative expansions.
So how can we combine the speed of \textsc{FORM} with the versatility and broadness of \textit{Mathematica}?
\item[]{\em Method of solution:} Named pipes, \textit{MathLink}, \textsc{C}, \textit{Mathematica} pre- and postprocessing.

\item[]{\em Typical running time:} Seconds, minutes, or more, depends on the
complexity of the calculation. There is some overhead for conversion of larger expressions.
\end{itemize}

\newpage

\hspace{1pc}
{\bf LONG WRITE-UP}

\section{B\lowercase{asic} I\lowercase{deas}\label{Basic Ideas}}
The basic and simplest way to communicate between \textit{Mathematica} and \textsc{FORM}~\cite{Kuipers:2012rf,Vermaseren:2000nd,Vermaseren:2011sb,Vermaseren:2010iw,Tentyukov:2008zz,Vermaseren:2008kw} is using input and output files.
This method has been used in \textsc{FeynCalc}~\cite{Mertig:1990an} and
\textsc{FormCalc} ~\cite{hep-ph/9807565},
which prepares the symbolic expressions of the diagrams in an input file for \textsc{FORM}, runs \textsc{FORM}, and retrieves the results back to \textit{Mathematica}.

Another method to exchange data between different processes is to use pipes.
The detailed usage of pipe communication between \textsc{FORM} and other external programs is described in \cite{Tentyukov:2006ys}.
In this article we implement a user-friendly communication between
\textit{Mathematica} and \text{FORM} using both possibilities.
We would like to note that \textsc{FormCalc} and \texttt{FormGet.tm},
mentioned at \url{http://www.feynarts.de/formcalc}, also use
\textit{MathLink} and pipes, but do not provide a general interface from
\textit{Mathematica} to \textsc{FORM}.

The remainder of this section is intended for programmers only and not needed to
understand how to use \textsc{FormLink}.

The basic idea is that we create two unnamed pipes, one is the read-only descriptor with file handler \verb=r#= ,
the other is the write-only descriptor with file handler \verb=w#=, then start
\textsc{FORM} with the {\tt pipe} option automatically through the
\textit{MathLink} executable \textsc{FormLink}
\begin{verbatim}
    form -pipe r#,w# init
\end{verbatim}
where {\tt init} refers to the \textsc{FORM} code \verb=init.frm= , which is discussed in the following.
When the pipe connection has been established successfully, \textsc{FORM} sends
its Process Identifer (PID) in ASCII decimal format with an appended newline character to the descriptor {\tt w\#} and
then \textsc{FORM} will wait for the answer from the descriptor {\tt r\#}.
The answer must be two comma-separated integers in ASCII decimal format followed by a newline character.
The first integer corresponds to the \textsc{FORM} PID while the second one is
the PID of parent process which started \textsc{FORM}.
If the answer is not obtained after some time-out, or if it is not correct, i.e. it is not a list of two integers or the first integer is not the \textsc{FORM} PID,
then \textsc{FORM} fails.
When the channel has been established successfully, \textsc{FORM} will run the
code in the {\tt init.frm} file, containing the following instructions:
\begin{verbatim}
    Off Statistics;
    #ifndef `PIPES_'
        #message "No pipes found";
        .end;
    #endif
    #if (`PIPES_' <= 0)
        #message "No pipes found";
        .end;
    #endif
    #procedure put(fmt, mexp)
        #toexternal `fmt', `mexp'
        #toexternal "#THE-END-MARK#"
    #endprocedure
    #setexternal `PIPE1_';
    #toexternal "OK"
    #fromexternal
    .end
\end{verbatim}
The core parts are the last two lines before the {\tt .end} instruction.
The first line, \verb=#toexternal "OK"=, sends the word \verb=OK= in ASCII
string format from \textsc{FORM} to \textsc{FormLink}, it confirms that the communication channel has been established successfully,
otherwise \textsc{FormLink} will treat it as failed.
The second line blocks \textsc{FORM} and waits for the code which
will be sent from \textit{Mathematica}.
When the {\tt prompt}\footnote{The default prompt defined in \textsc{FORM} is a blank line, for details please see \cite{Tentyukov:2006ys}}
arrives,
\textsc{FORM} will continue to execute code from \verb=#fromexternal=.
We defined a procedure named {\tt put} to send data from \textsc{FORM} back to \textit{Mathematica}.
Note that if you want to send data without this procedure, you need to send the end mark,
i.e. the string \verb=#THE-END-MARK#=, to indicate that the data is complete,
otherwise \textsc{FormLink} will be blocked until the end mark has been received.

Until now we have demonstrated a round communication from \textit{Mathematica} to \textsc{FORM}, and then back to \textit{Mathematica} again.
This procedure can be looped if we put another \verb=#fromexternal= in the code sent from \textit{Mathematica} to \textsc{FORM},
and in this sense, we get an interactive \textsc{FORM},
which cannot be easily achieved by the first way which exchanges data by input and output files.

\section{I\lowercase{nstallation}}
To install the \textsc{FormLink} and \textsc{FeynCalcFormLink} packages, run the
following instruction in a Kernel or Notebook session of \textit{Mathematica} 7,
8 or 9 on Linux, MacOSX, or Windows.
\begin{verbatim}
    Import["http://www.feyncalc.org/formlink/install.m"]
\end{verbatim}
The installer will automatically download \texttt{formlink.zip} from the url\footnote{\url{http://www.feyncalc.org/formlink/formlink.zip}}
and extract the files from the archive to the directory \texttt{Applications} in the directory \texttt{\$UserBaseDirectory},
which is by default located in the search path of \textit{Mathematica}.
Their values for different platform are listed in TABLE~\ref{userbasedirectory}.
Specifying
\begin{verbatim}
    $installdirectory = mydir
\end{verbatim}
before running the installer will change the installation directory. It is
recommended to use a directory which is on the \textit{Mathematica} path, e.g.,
 \texttt{HomeDirectory[]}, or \texttt{\$BaseDirectory}.

For user convenience, the binary files of {\tt FormLink}, {\tt form} and {\tt tform} for Linux, Microsoft Windows and
Mac OS X are also installed.
Furthermore the latest version of {\sc FeynCalc} is downloaded from \url{http://www.feyncalc.org} and installed automatically into the same directory
unless it has been already installed somewhere on the \textit{Mathematica} path.
\begin{table}[htbp]
\begin{tabularx}{0.9\textwidth}{cC{1cm}X}
\toprule
Platform & & {\tt \$UserBaseDirectory} \\
\midrule
{\tt Windows} & & {\tt
C:\textbackslash{}Users\textbackslash{}{\it username}\textbackslash{}AppData\textbackslash{}Roaming\textbackslash{}Mathematica}\\
{\tt Linux} && {\tt \textasciitilde/.Mathematica} \\
{\tt Mac OS X} & & {\tt \textasciitilde/Library/Mathematica} \\
\bottomrule
\end{tabularx}
\caption{The values of {\tt \$UserBaseDirectory} for different operating systems\label{userbasedirectory}}
\end{table}

At the end of the installation {\tt FormLink} and {\tt FeynCalcFormLink} are loaded and
two simple examples are run, one uses \texttt{FormLink}:

\vspace{.3cm}\noindent\includegraphics{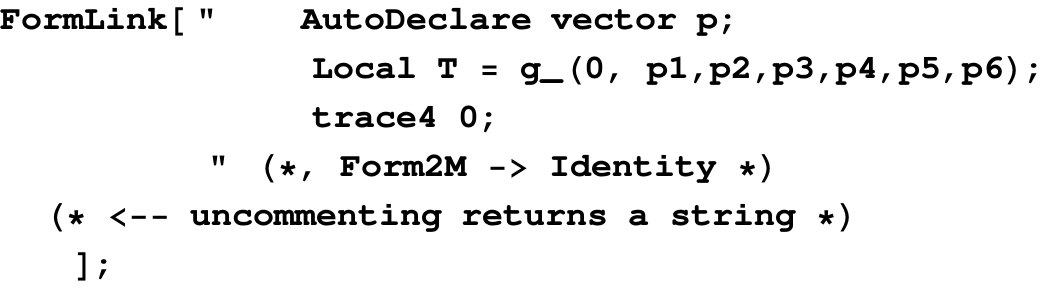}

\vspace{.3cm}\noindent\includegraphics{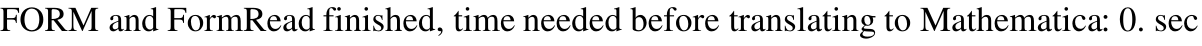}

\vspace{.03cm}\noindent\includegraphics{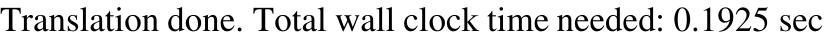}

\vspace{.03cm}\noindent\includegraphics{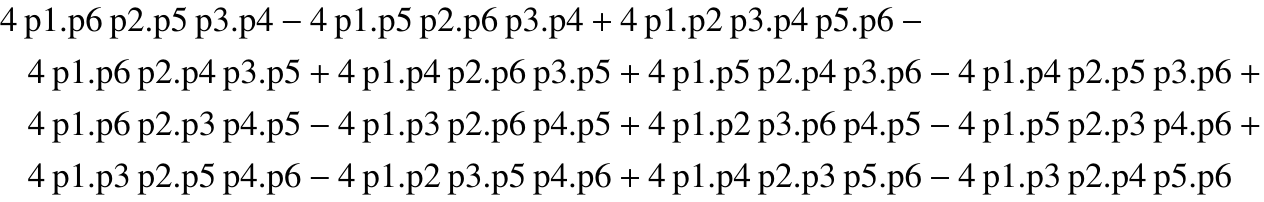}

\vspace{0.5cm}
\noindent The other example executes \texttt{FeynCalcFormLink}, which generates
and runs the corresponding \textsc{FORM}  program, substituting ASCII values for
greek indices intermediately:

\vspace{.3cm}\noindent\includegraphics{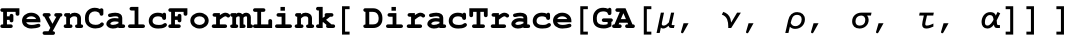}

\vspace{-0.cm}\noindent\rule{\textwidth}{0.1pt}\\{\color{Code}
\texttt{\sf AutoDeclare Index lor;\\
Format Mathematica;\\
L resFL = (g$\_$(1,lor2)*g$\_$(1,lor3)*g$\_$(1,lor4)*g$\_$(1,lor5)*g$\_$(1,lor6)*g$\_$(1,lor1));\\
trace4,1;\\
contract 0;\\
.sort;\\
$\#$call put({``}$\%$E{''}, resFL)\\
$\#$fromexternal}}

\noindent\rule[.4\baselineskip]{\textwidth}{0.1pt}

\vspace{.03cm}\noindent\includegraphics{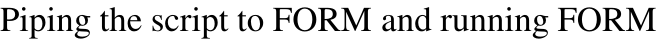}

\vspace{.03cm}\noindent\includegraphics{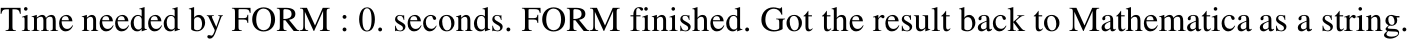}

\vspace{.03cm}\noindent\includegraphics{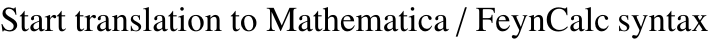}

\vspace{.03cm}\noindent\includegraphics{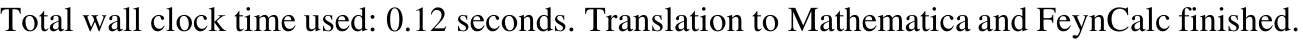}

\vspace{.03cm}\noindent\includegraphics{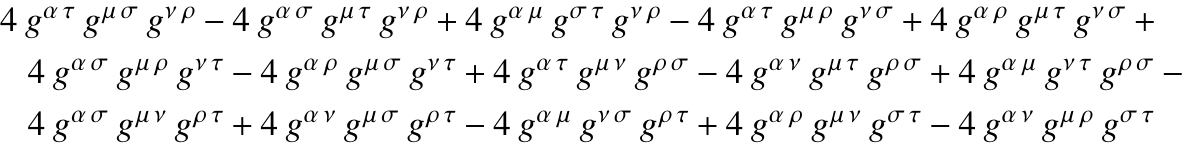}

\vspace{.5cm}
\begin{table}[!th]
\begin{tabularx}{\textwidth}{cC{.5cm}X}
\toprule
Platform && {\tt make} commands \\
\midrule
\multirow{2}{*}{\tt Microsoft Windows (32 \& 64 bit)} && {\tt make -f makefile.cygwin } \\
&& {\tt make -f makefile.cygwin install} \\ \hline
\multirow{2}{*}{\tt Linux (32 bit)} && {\tt make -f makefile.linux32 } \\
&& {\tt make -f makefile.linux32 install} \\ \hline
\multirow{2}{*}{\tt Linux (64 bit)} && {\tt make -f makefile.linux64 } \\
&& {\tt make -f makefile.linux64 install} \\ \hline
\multirow{2}{*}{\tt Mac OS X (64 bit)} && {\tt make -f makefile.macosx64 } \\
&& {\tt make -f makefile.macosx64 install} \\
\bottomrule
\end{tabularx}
\caption{The {\tt make} commands for different operating systems.
If needed, please modify the corresponding makefile, for example, change the location of
the \textit{MathLink} developer kit ({\tt MLINKDIR}) if you did not use the
default location during installation of \textit{Mathematica} or if you want to use a different version than \textit{Mathematica} 8.
\label{makecommands}
}
\end{table}

The source code of {\sc FormLink} is included in the \texttt{src} folder.
The scripts for compilation are listed in TABLE~\ref{makecommands}.
We used the GNU compiler \texttt{gcc}, producing an executable \texttt{FormLink}
in the corresponding subdirectory in \texttt{bin}.

\section{F\lowercase{orm}L\lowercase{ink and} F\lowercase{eyn}C\lowercase{alc}F\lowercase{orm}L\lowercase{ink} F\lowercase{unctions} R\lowercase{eference}}
\subsection{Basic Usage}
The basic functions are \texttt{FormLink} to execute \textsc{FORM} programs
written as a string in \textit{Mathematica}
and \texttt{FeynCalcFormLink} to run a program specified in \textsc{FeynCalc} syntax through \textsc{FORM}.
These two \textit{MathLink}-based functions are easy to use for \textsc{FORM} and \textsc{FeynCalc} practicioners.
\begin{itemize}
\item \verb=FormLink=\par
\verb=FormLink[" form statements "]= runs the {\tt form statements} in \textsc{FORM}
and returns the result to \textit{Mathematica}. If only one \texttt{Local}
assignment is present the return value is the result.
If more {\sc FORM} \texttt{Local} variables are present, a list of
results is returned. \verb=FormLink= has several options, the default values are as follows,\par \includegraphics{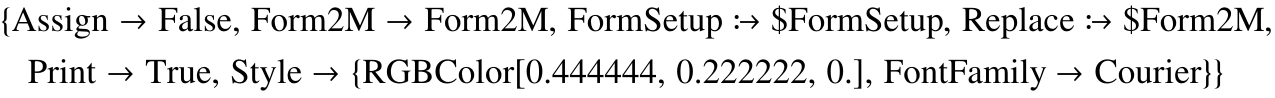}\par
Here we only explain the most important ones.

The option \texttt{Form2M} can be used to specify the function which translates
the expression from \textsc{FORM} to \textit{Mathematica} format, its default
value is function \texttt{Form2M}, which has the following usage: {\tt Form2M[string, replist]} translates {\tt string} by {\tt ToExpression[StringReplace[string,replist],TraditionalForm]} to {\it Mathematica}.
\texttt{Form2M} can also be set to {\tt Identity}, which means that the result
from \textsc{FORM} will be returned as a string, containing the exact output
for the \texttt{Local} expression.
If conversion to \textit{Mathematica} is straighforward, i.e., there are no
bracketed expressions or other syntaxes not easily interpreted by
\textit{Mathematica}, then it is best to use the setting \texttt{Form2M$\to$ToExpression},
which will just call \texttt{ToExpression} on the string result coming from \textsc{FORM}.

The option \texttt{Replace} is a list of string replacements, its default
setting is \texttt{\$Form2M}, containing a list of user-changeable (e.g. in Config.m) basic function-name translations from \textsc{FORM} to \textit{Mathematica}.\par
\vspace{.03cm}\noindent\includegraphics{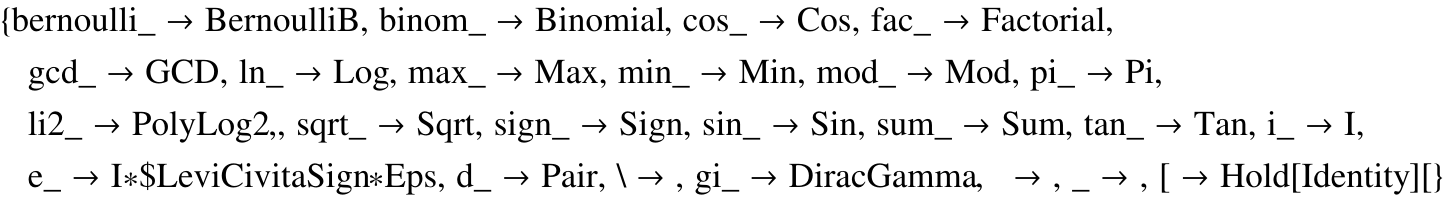}

The option \texttt{Print} can be used to switch informative messages on or off during execution.
The option  \texttt{FormSetup} can be a list of \textsc{FORM} settings, like \texttt{TempDir}
which dynamically produces a \texttt{form.set} file next to the binary which is then called automatically upon starting \textsc{FORM}.
\end{itemize}

\begin{itemize}
\item \verb=FeynCalcFormLink=\par
\verb=FeynCalcFormLink[expr]= translates the \textsc{FeynCalc} expression
\texttt{expr} to a \textsc{FORM} program, identifying automatically traces,
symbols, vectors, indices and the dimension,
calculates it, pipes it back to \textit{Mathematica} and translates it to \textsc{FeynCalc} syntax.

There are several options for \verb=FeynCalcFormLink=, the default settings are
as follows,\par \includegraphics{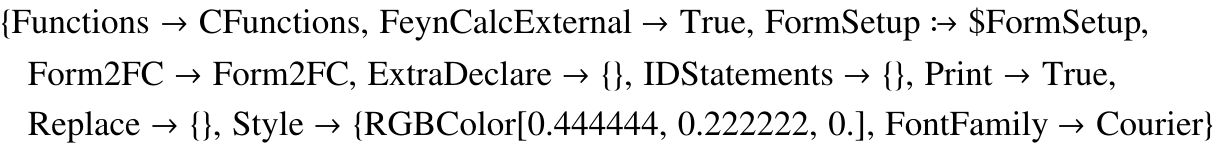}\par
Most of the options are similar to \texttt{FormLink}, we only explain
{\tt Functions}, \texttt{ExtraDeclare}, \texttt{IDStatements} and {\tt Form2FC}.

If the option {\tt Functions} is set to {\tt ``CFunctions''}, then all non-{\tt
System`} functions, except those present in {\tt \$M2Form} and some \textsc{FeynCalc} functions, are automatically declared {\tt CFunctions} in FORM.
If {\tt Functions} option is set to {\tt ``Functions''}, then they are declared
noncommutative functions, i.e.,  {\tt Functions} in \textsc{FORM}.

The option {\tt ExtraDeclare} can be used to put any extra valid \textsc{FORM}
declarations which are not identified automatically,  for example:\par
\texttt{ExtraDeclare$\to$\{"CFunctions GammaFunction;", "Functions MyOperator;"\}}

The option \texttt{IDStatements} can be set to a string or a list of strings corresponding to \textsc{FORM} \texttt{identify} statement like \verb+{"id k1.k1=mass^2;"}+.

The option {\tt Form2FC} is set to the function being used to translate
expression from \textsc{FORM} to \textsc{FeynCalc} format, the default is {\tt Form2FC}.
\end{itemize}

\subsection{Advanced Usage}

The basic internal procedure to use \textsc{FormLink} is to start \textsc{FORM} with \texttt{FormStart},
then send the code to \textsc{FORM} for execution using \texttt{FormWrite} and \texttt{FormPrompt},
read and convert the result back to \textit{Mathematica} through \texttt{FormRead}, finally, stop \textsc{FORM} by calling \texttt{FormStop}.
All these calls can be encoded into a single function named \texttt{FormLink} which has been introduced in the previous section.

\begin{itemize}

\item \verb=FormStart=\par
\verb=FormStart[]= automatically determines the path of the \texttt{form}
executable, which is located in the corresponding subfolder in the \texttt{bin} directory,
and then starts \textsc{FORM} in pipe mode.
You can also call \verb=FormStart[formpath]= with explicit full path
\texttt{formpath} of the \texttt{form} excutable.

\item \verb=FormWrite=\par
\verb=FormWrite[script]= sends the \texttt{script} with an appended newline character, to \textsc{FORM}.
Notice that \textsc{FORM} will not start to execute the script right away
and you can send your scripts by using the \verb=FormWrite[script]= many times.
When you are finished, send the prompt, then \textsc{FORM} executes all instructions sent.
\par \verb=FormWrite[scripts]=, where \texttt{scripts} is a list of strings, will call \verb=FormWrite[script]= for each element \texttt{script} in
the list scripts, so if your code spans several lines, you can also put them as a list with each element corresponding to a single line of your scripts.

\item \verb=FormPrompt=\par
\verb=FormPrompt[]= sends the prompt to \textsc{FORM}, which will make \textsc{FORM} continue to execute the code you have sent.
The default prompt in \textsc{FORM} is a blank line, \textsc{FormLink} has adopted this default option, so do not send blank lines unintentionally.

\item \verb=FormRead=\par
\verb=FormRead[]= reads the data from \textsc{FORM}. It should be noted that if no data is sent from \textsc{FORM}, the calling thread will be blocked until data is available.
You can use the procedure \verb=put= defined in \verb=init.frm= to send the data from \textsc{FORM}. \par
\verb=FormRead[]= will first check whether the pipe has been closed or not, if the pipe has been closed, for example when \textsc{FORM} has encountered some problems,
\verb=FormRead[]= will redirect the standard output from \textsc{FORM} to \textit{Mathematica}, so the user can check the error messages.

\item \verb=FormStop=\par
\verb=FormStop[]= uninstalls the link to Form. \verb=FormStop[All]= kills all running FormLink processes.
\end{itemize}

As we discussed in section~\ref{Basic Ideas}, there are two ways to communicate between \textsc{FORM} and \textit{Mathematica},
the functions introduced above are all used with the method of piping.
We also provide two functions which are not using \textit{MathLink}, but deal with input and output files only.

\begin{itemize}
\item \verb=RunForm=\par
{\tt RunForm[script]} runs {\tt script} in \textsc{FORM} and writes the result
to {\tt runform.frm} and the log file to {\tt form.log} in the current
directory, i.e., the value returned by the {\textit Mathematica} function {\tt Directory[]}.

{\tt RunForm[script, formfile]} uses {\tt formfile} instead of {\tt runform.frm}. The first argument {\tt script} can be a string or a list of strings.

An optional third argument can be given to use a specific {\sc FORM} executable, otherwise a {\sc FORM} executable from {\tt \$FormLinkDir/bin} is used.
\item \verb=ReadString=\par
{\tt ReadString[str]} imports {\tt str} as {\it Text} and translates it to {\it Mathematica} syntax by using {\tt \$Form2M}.
You can use \verb=#write= preprocessor to write your data to the output file
by \textsc{FORM}, and use {\tt ReadString} to read it into \textit{Mathematica}.
\end{itemize}

To facilitate the usage of \textsc{FormLink} with \textsc{FeynCalc}, two functions are provided for performing the conversions between \textsc{FeynCalc} and \textsc{FORM}\footnote{There are also two older \textsc{FeynCalc} functions named {\tt FeynCalc2FORM} and {\tt FORM2FeynCalc} to perform the conversions.}:
\begin{itemize}
\item \verb=FC2Form=\par
\verb=FC2Form[exp]= translates {\tt exp} in \textsc{FeynCalc} format to \textsc{FORM} format.
\verb=FC2Form[exp]= returns a list of two elements, the first one
is a list containing the script which will be sent to \textsc{FORM}.
You can use the function \verb=ShowScript[script]= to display the {\tt script}.
The second one is  a list of replacement rules, which will be used in
\verb=Form2FC= to translate the result from \textsc{FORM} format back to \textsc{FeynCalc}.
\verb=FC2Form= has the options : {\tt Functions}, {\tt Dimension}, {\tt ExtraDeclare}, {\tt IDStatements}, {\tt Print} and {\tt Replace}:

\vspace{.03cm}\noindent\includegraphics{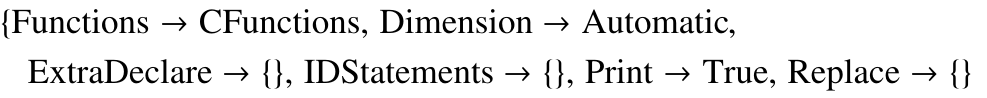}\par

\item \verb=Form2FC=\par
\verb=Form2FC[formexpr]= is used to translate {\tt formexpr} which is in
\textsc{FORM} format back to the \textsc{FeynCalc} one.
\verb=Form2FC[formexpr, replacelist]= applies the substitution list
{\tt replacelist} at the end. The second argument is usually the second item of
of the list returned by \verb=FC2Form=.
\end{itemize}

So the general steps to use the package with \textsc{FeynCalc} are to convert \textsc{FeynCalc} code to \textsc{FORM}, and then send the converted code to \textsc{FORM} for executing
with \textsc{FormLink}, and finally convert the results in \textsc{FORM} format
back to \textsc{FeynCalc}.
All these steps are implemented into a single function {\tt FeynCalcFormLink},
which has been introduced in the previoius section.

\section{E\lowercase{xamples} U\lowercase{sing} F\lowercase{orm}L\lowercase{ink}, F\lowercase{eyn}C\lowercase{alc}F\lowercase{orm}L\lowercase{ink and }R\lowercase{un}F\lowercase{orm}}
We list a few examples using the {\tt FormLink}, {\tt FeynCalcFormLink} and {\tt
RunForm} functions.
More examples can be found in the {\tt Examples} directories of both packages.
\subsection{Using \texttt{FormLink}}

{\tt FormLink} is a function for running {\sc FORM} from \textit{Mathematica} and returning the result to \textit{Mathematica}.

\subsec{A short trace}

\vspace{.03cm}\noindent\includegraphics{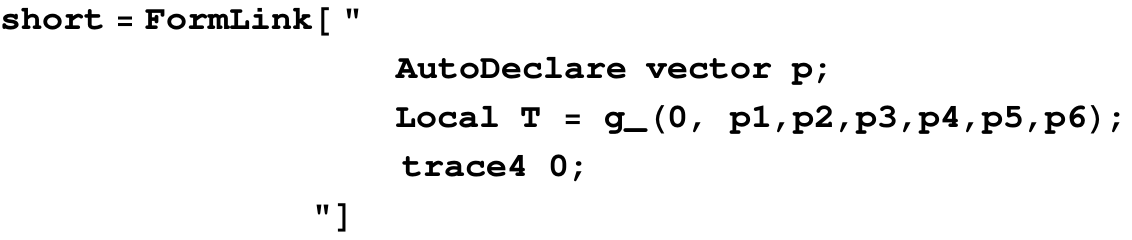}

\vspace{.03cm}\noindent\includegraphics{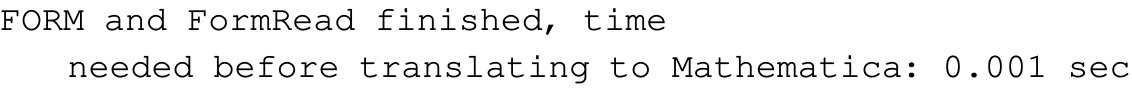}

\vspace{.03cm}\noindent\includegraphics{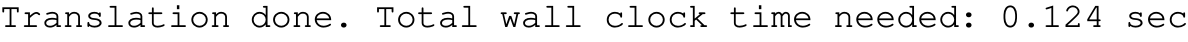}

\vspace{.03cm}\noindent\includegraphics{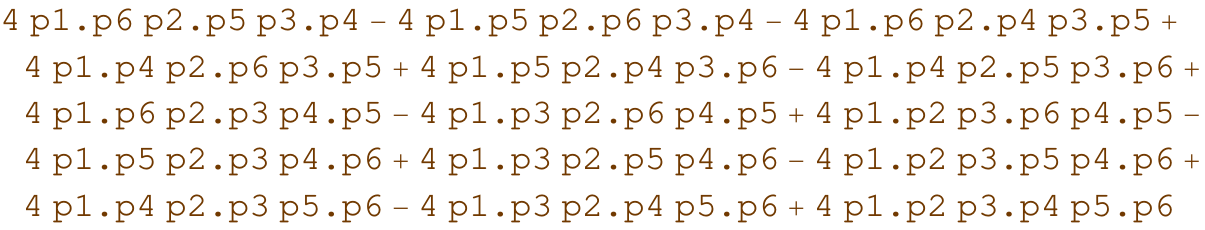}

\vspace{.03cm}\noindent\includegraphics{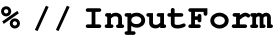}

\vspace{.03cm}\noindent\includegraphics{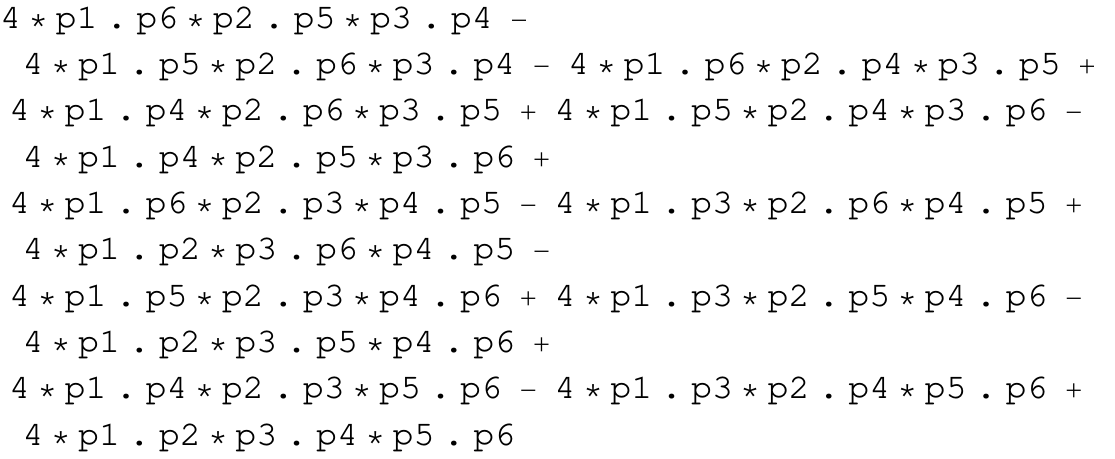}

\vspace{.5cm}

\noindent Using the option {\tt Form2M$\to$Identity} returns a string:

\vspace{.5cm}

\vspace{.03cm}\noindent\includegraphics{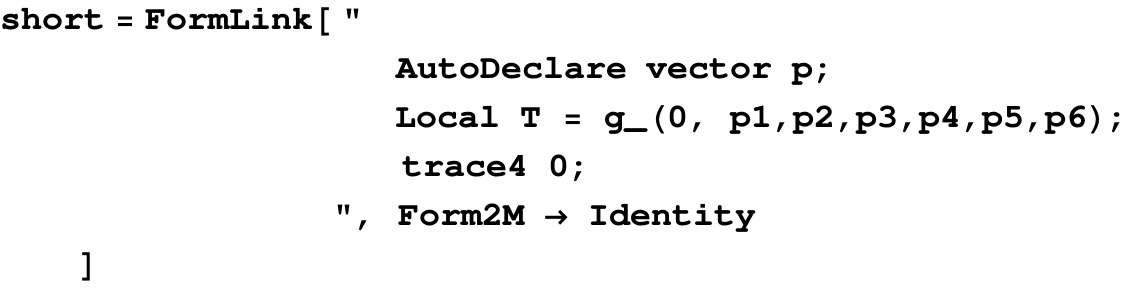}

\vspace{.03cm}\noindent\includegraphics{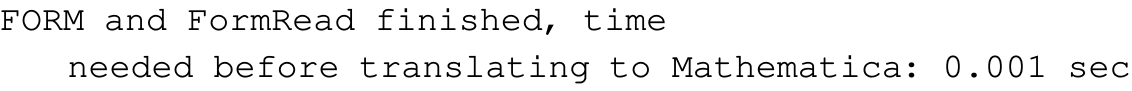}

\vspace{.03cm}\noindent\includegraphics{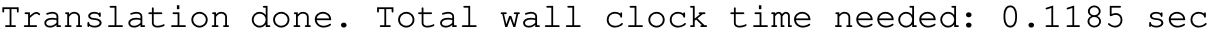}

\vspace{.03cm}\noindent\includegraphics{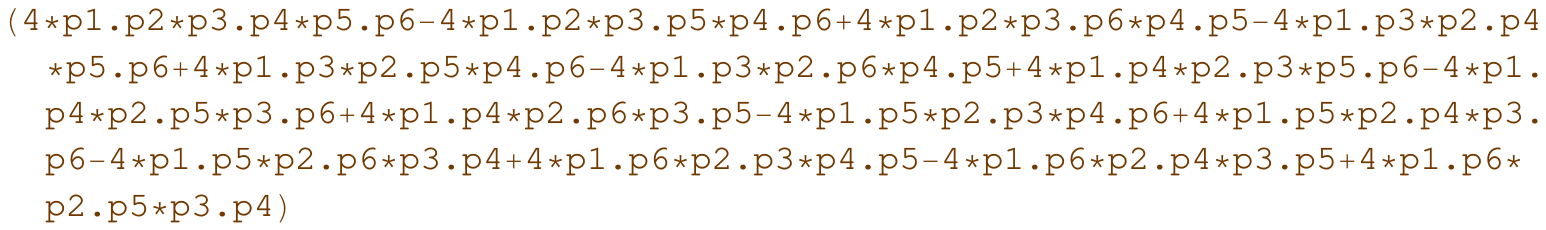}

\vspace{.03cm}\noindent\includegraphics{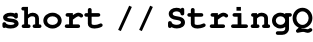}

\vspace{.03cm}\noindent\includegraphics{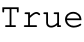}

\vspace{.03cm}\noindent\includegraphics{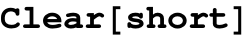}

\newpage
\subsec{A longer trace}

\vspace{.03cm}\noindent\includegraphics{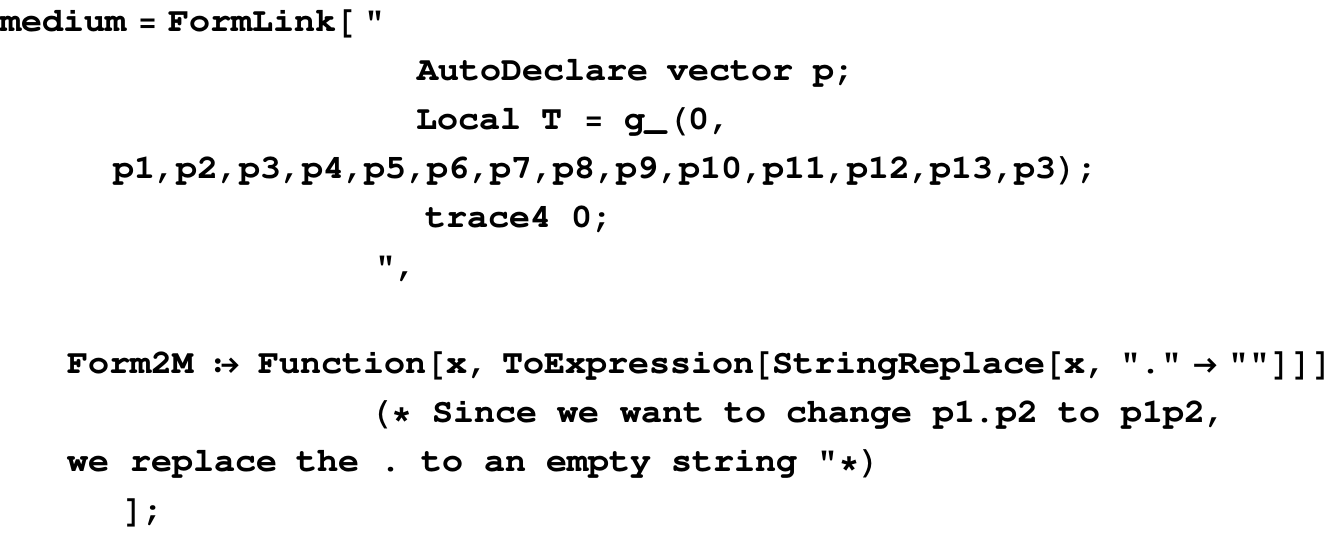}

\vspace{.03cm}\noindent\includegraphics{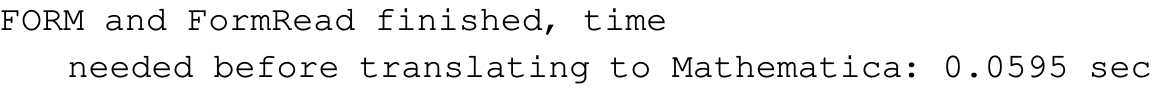}

\vspace{.03cm}\noindent\includegraphics{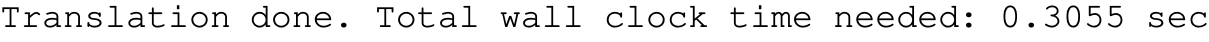}

\vspace{.03cm}\noindent\includegraphics{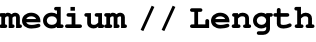}

\vspace{.03cm}\noindent\includegraphics{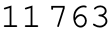}

\vspace{.03cm}\noindent\includegraphics{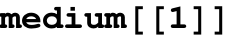}

\vspace{.03cm}\noindent\includegraphics{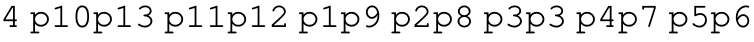}

\vspace{.03cm}\noindent\includegraphics{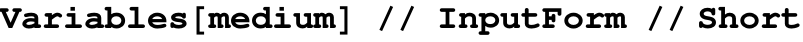}

\vspace{.03cm}\noindent\includegraphics{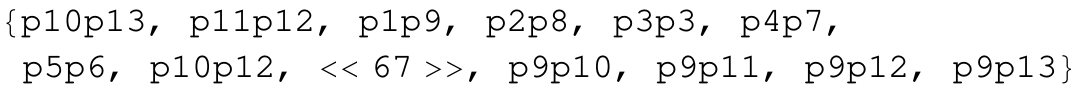}

\vspace{.03cm}\noindent\includegraphics{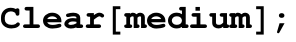}

\subsec{Special cases }

\noindent Objects like {\tt [1-x]} in {\sc FORM} are translated to {\tt (1-x)} in \textit{Mathematica}:

\vspace{0.5cm}
\vspace{.03cm}\noindent\includegraphics{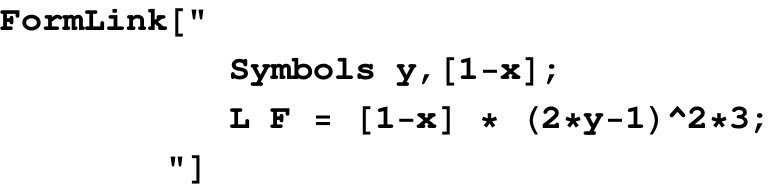}

\vspace{.03cm}\noindent\includegraphics{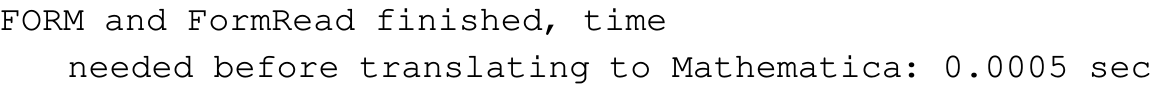}

\vspace{.03cm}\noindent\includegraphics{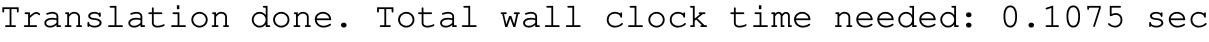}

\vspace{.03cm}\noindent\includegraphics{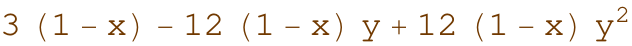}

\vspace{.03cm}\noindent\includegraphics{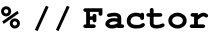}

\vspace{.03cm}\noindent\includegraphics{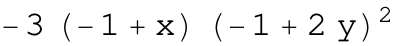}

\vspace{.03cm}\noindent\includegraphics{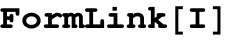}

\vspace{.03cm}\noindent\includegraphics{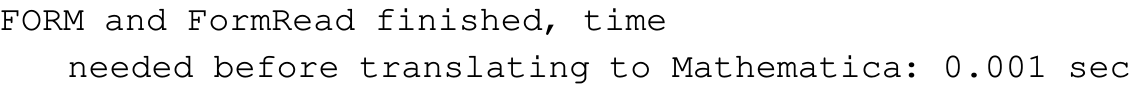}

\vspace{.03cm}\noindent\includegraphics{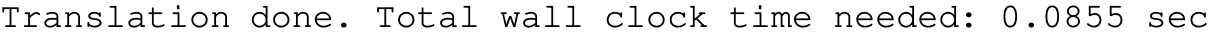}

\vspace{.03cm}\noindent\includegraphics{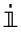}

\newpage
\subsec{Polynomial examples}

\subsec{Expand $(a+b)^2$}

\vspace{.03cm}\noindent\includegraphics{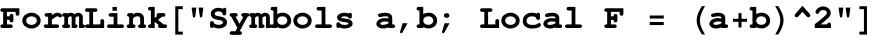}

\vspace{.03cm}\noindent\includegraphics{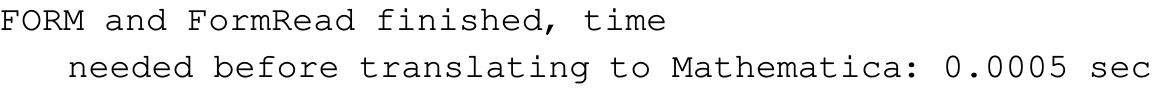}

\vspace{.03cm}\noindent\includegraphics{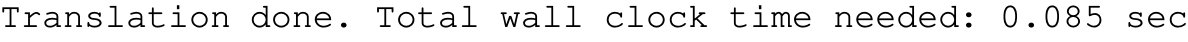}

\vspace{.03cm}\noindent\includegraphics{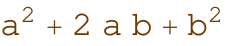}

\vspace{0.5cm}
\noindent Alternatively we may just enter the \textit{Mathematica} expression which is then automatically translated to a {\sc FORM} program:
\vspace{0.5cm}

\vspace{.03cm}\noindent\includegraphics{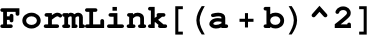}

\vspace{.03cm}\noindent\includegraphics{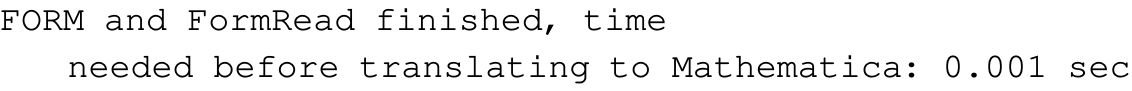}

\vspace{.03cm}\noindent\includegraphics{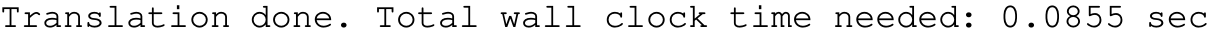}

\vspace{.03cm}\noindent\includegraphics{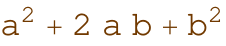}

\subsec{Expand $(a+b+c+d+e+f+g)^{21}$}

\vspace{0.5cm}
\noindent The default option setting of {\tt Form2M} does general translation of {\sc FORM} syntax to \textit{Mathematica}/{\sc FeynCalc}. Here we can use the simpler {\tt ToExpression} as a setting of {\tt Form2M}.

\noindent While it is possible to use even 42 instead of 21, we see already for power 21 that it takes longer to transfer the large expression between {\sc FORM} and \textit{Mathematica} than to calculate it in either FORM or \textit{Mathematica} directly. So it is best not to transfer large expressions, if possible.
\vspace{0.5cm}

\vspace{.03cm}\noindent\includegraphics{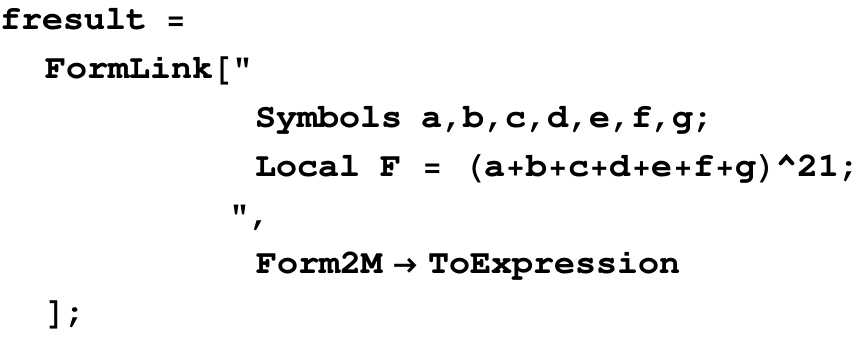}

\vspace{.03cm}\noindent\includegraphics{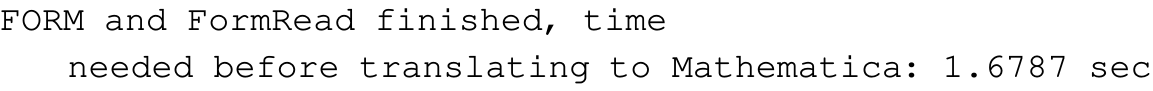}

\vspace{.03cm}\noindent\includegraphics{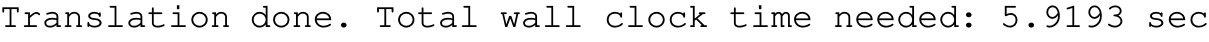}

\vspace{.03cm}\noindent\includegraphics{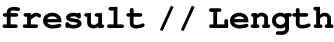}

\vspace{.03cm}\noindent\includegraphics{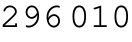}

\vspace{0.5cm}
\noindent So for somewhat larger expressions there is noticable overhead for piping the expression back to \textit{Mathematica} as a string by {\tt FormRead} and for translating the string to \textit{Mathematica} syntax ({\tt ToExpression}).

\noindent An accurate timing of {\sc FORM} can be done by using {\tt RunForm}:
\vspace{0.5cm}

\vspace{.03cm}\noindent\includegraphics{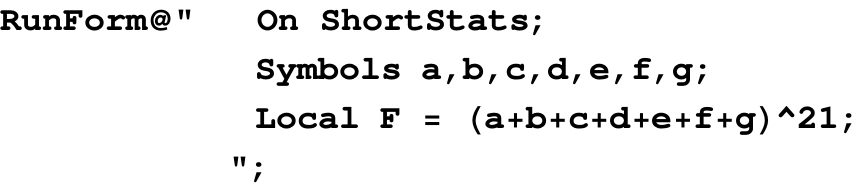}

\vspace{.03cm}\noindent\includegraphics{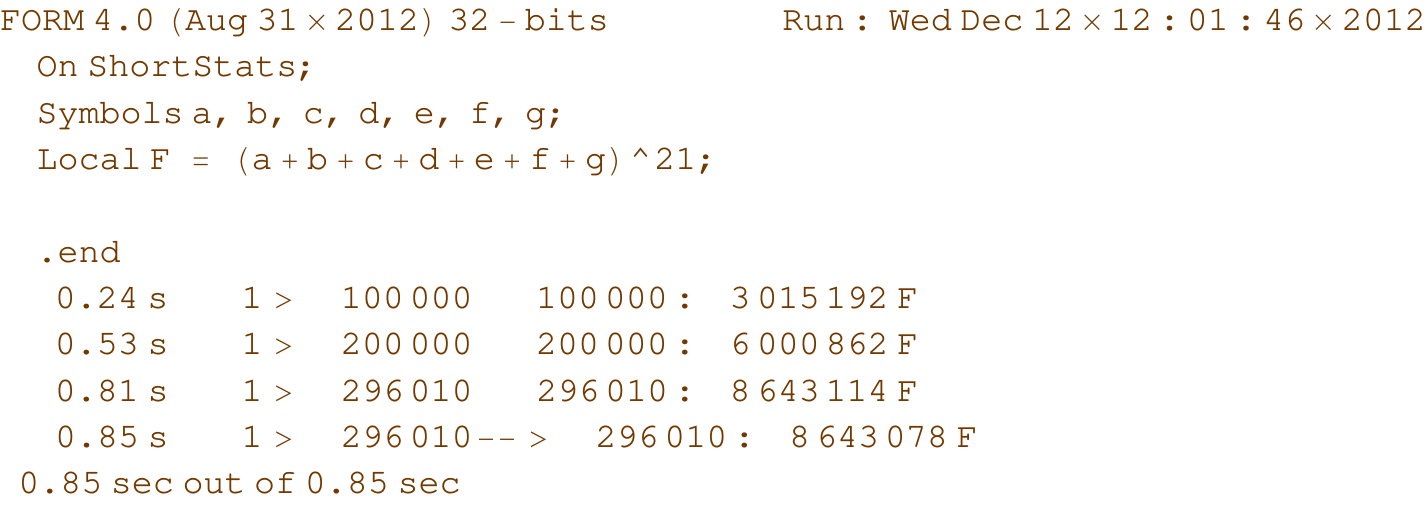}

\subsec{Expand $(a+b+c+d+e+f+g)^{42}$ using {\tt RunForm}}


\noindent {\tt Last[StringSplit[$\#$,{''}$\backslash $n{''}]]$\&$} is a pure function to extract the last line of the string output of {\tt RunForm}, returning the timing.
\vspace{0.5cm}

\vspace{.03cm}\noindent\includegraphics{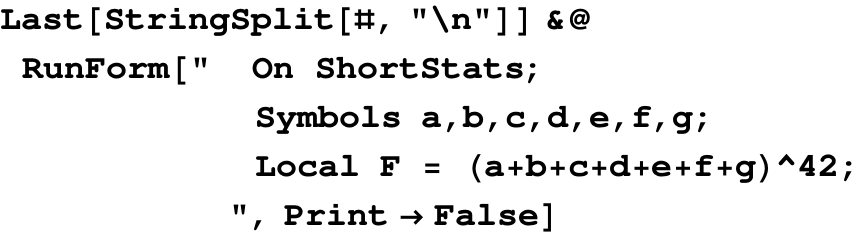}

\vspace{.03cm}\noindent\includegraphics{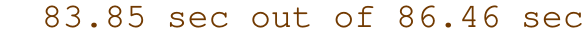}

\vspace{.03cm}\noindent\includegraphics{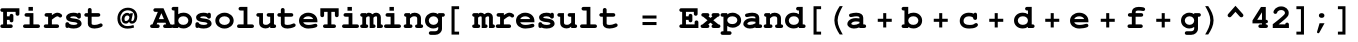}

\vspace{.03cm}\noindent\includegraphics{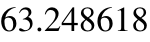}

\vspace{.03cm}\noindent\includegraphics{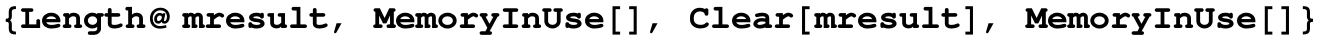}

\vspace{.03cm}\noindent\includegraphics{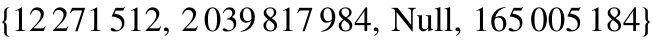}

\vspace{0.5cm}
\noindent So {\tt mresult} has more than 12 million terms and it occupies around 2 GB.
For this example with a lot of terms \textsc{FORM 4} is slightly slower than \textit{Mathematica} 9.
\vspace{0.5cm}

\vspace{.03cm}\noindent\includegraphics{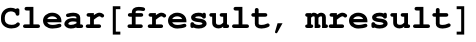}

\subsec{$(a+b)^6$ and $a^2 b\to c$}

\vspace{.03cm}\noindent\includegraphics{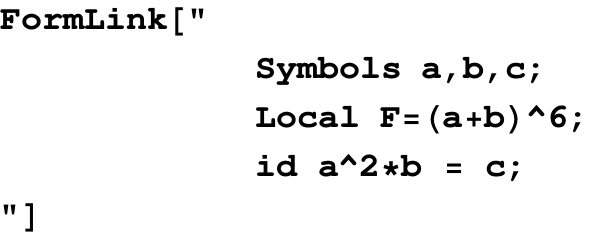}

\vspace{.03cm}\noindent\includegraphics{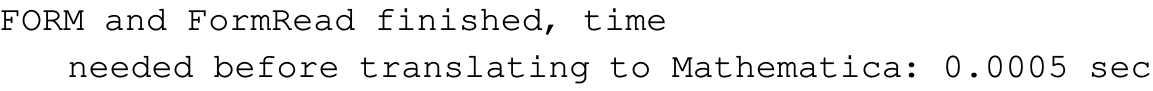}

\vspace{.03cm}\noindent\includegraphics{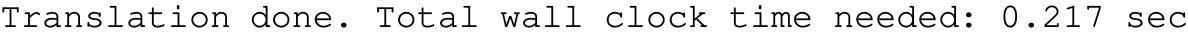}

\vspace{.03cm}\noindent\includegraphics{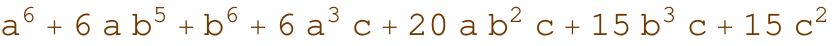}

\noindent To do the same in \textit{Mathematica} requires more work:
\vspace{0.5cm}

\vspace{.03cm}\noindent\includegraphics{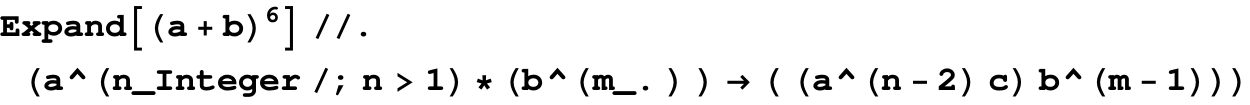}

\vspace{.03cm}\noindent\includegraphics{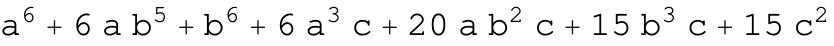}

\subsec{Multiple Local variables: Local F1= $(a+b+c)^{10}$ \ \  Local F2 = $(a+b+c+d)^{10}$}

\vspace{.03cm}\noindent\includegraphics{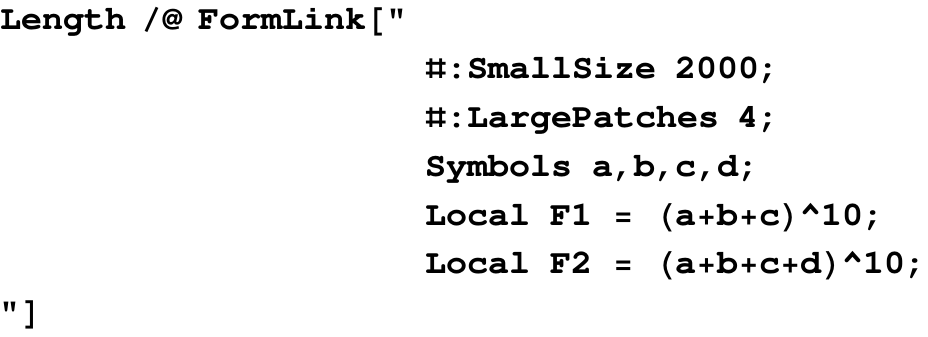}

\vspace{.03cm}\noindent\includegraphics{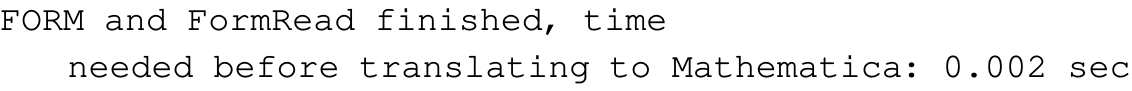}

\vspace{.03cm}\noindent\includegraphics{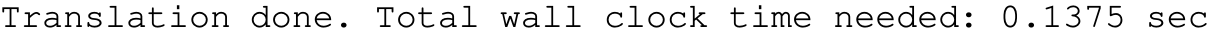}

\vspace{.03cm}\noindent\includegraphics{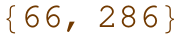}

\vspace{.03cm}\noindent\includegraphics{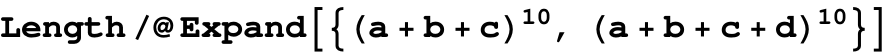}

\vspace{.03cm}\noindent\includegraphics{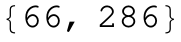}

\subsec{Commuting and noncommuting functions}

\vspace{.03cm}\noindent\includegraphics{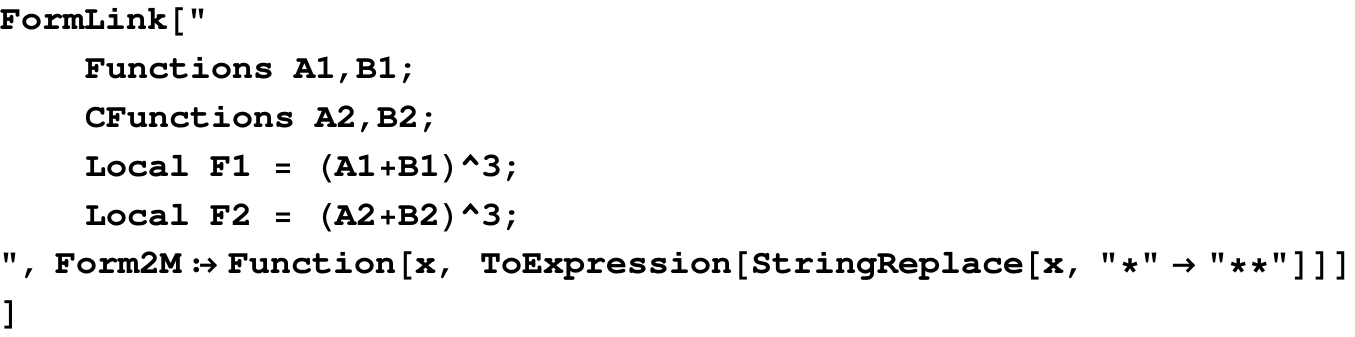}

\vspace{.03cm}\noindent\includegraphics{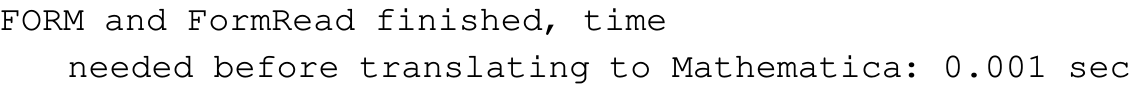}

\vspace{.03cm}\noindent\includegraphics{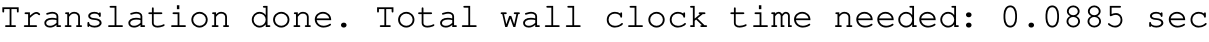}

\vspace{.03cm}\noindent\includegraphics{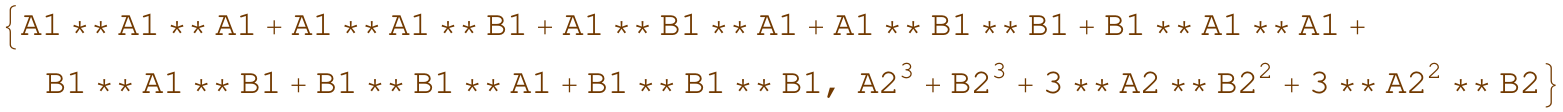}

\subsec{Index and Vector (Local F=p1(i1)*(p2(i1)+p3(i3))*(p1(i2)+p2(i3));)}

\vspace{.03cm}\noindent\includegraphics{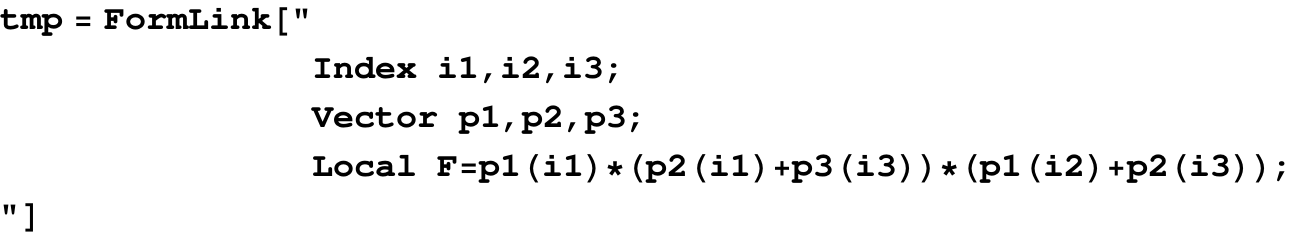}

\vspace{.03cm}\noindent\includegraphics{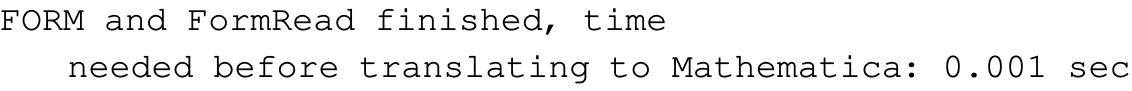}

\vspace{.03cm}\noindent\includegraphics{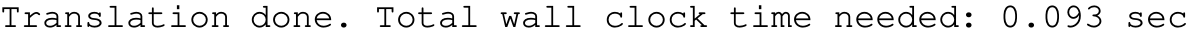}

\vspace{.03cm}\noindent\includegraphics{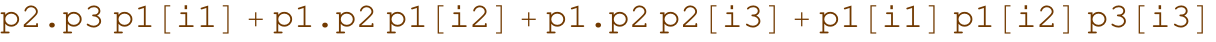}

\vspace{.03cm}\noindent\includegraphics{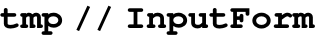}

\vspace{.03cm}\noindent\includegraphics{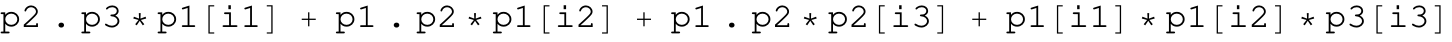}

\vspace{.03cm}\noindent\includegraphics{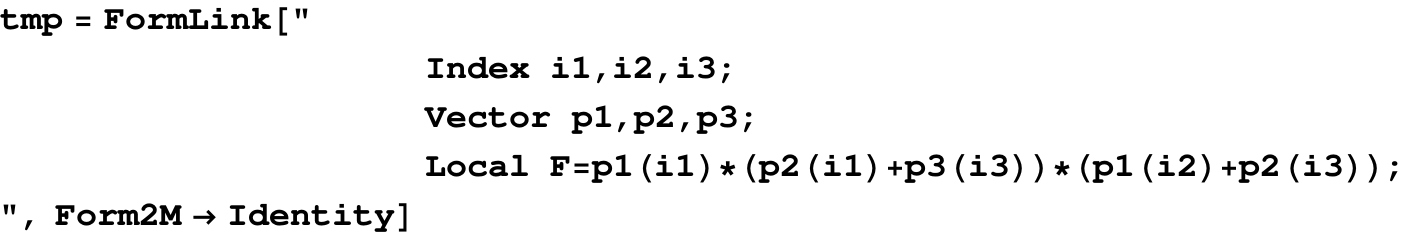}

\vspace{.03cm}\noindent\includegraphics{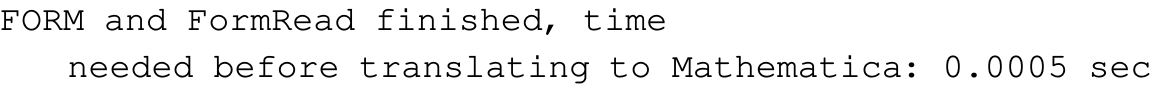}

\vspace{.03cm}\noindent\includegraphics{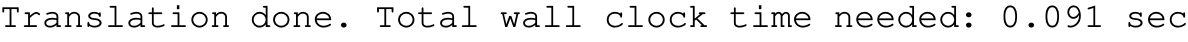}

\vspace{.03cm}\noindent\includegraphics{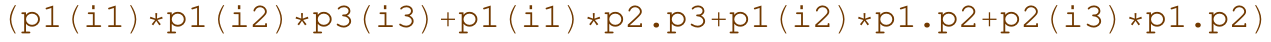}

\vspace{.03cm}\noindent\includegraphics{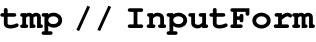}

\vspace{.03cm}\noindent\includegraphics{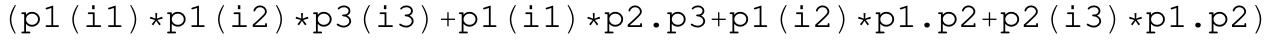}

\vspace{.03cm}\noindent\includegraphics{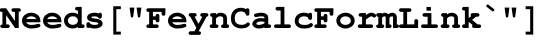}

\vspace{.03cm}\noindent\includegraphics{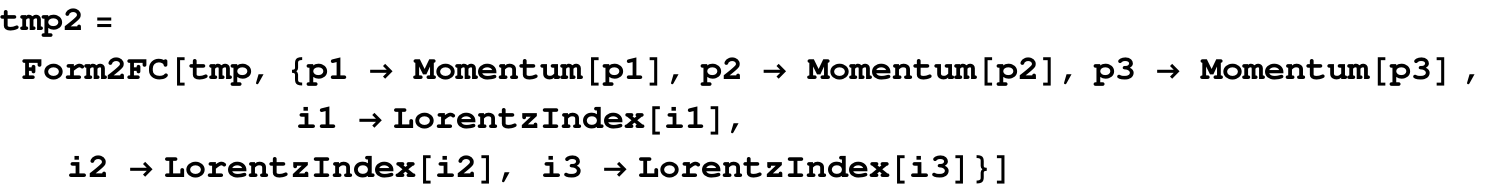}

\vspace{.03cm}\noindent\includegraphics{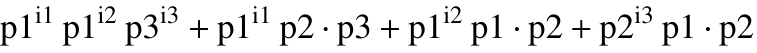}

\vspace{.03cm}\noindent\includegraphics{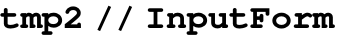}

\vspace{.03cm}\noindent\includegraphics{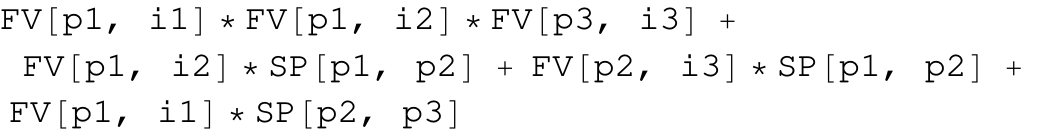}

\vspace{.03cm}\noindent\includegraphics{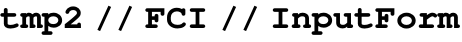}

\vspace{.03cm}\noindent\includegraphics{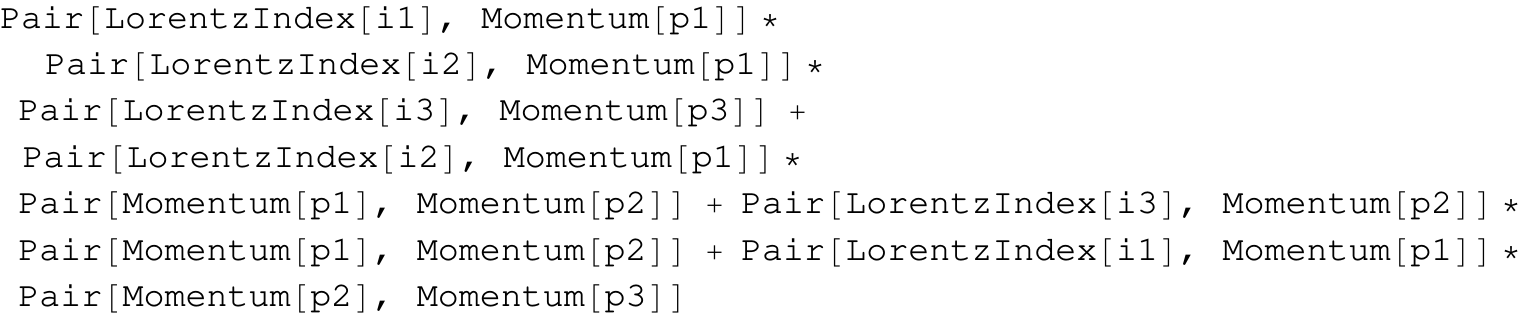}

\vspace{.03cm}\noindent\includegraphics{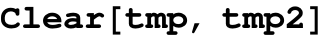}

\noindent This sets back the default output format type to {\tt StandardForm}.

\vspace{.03cm}\noindent\includegraphics{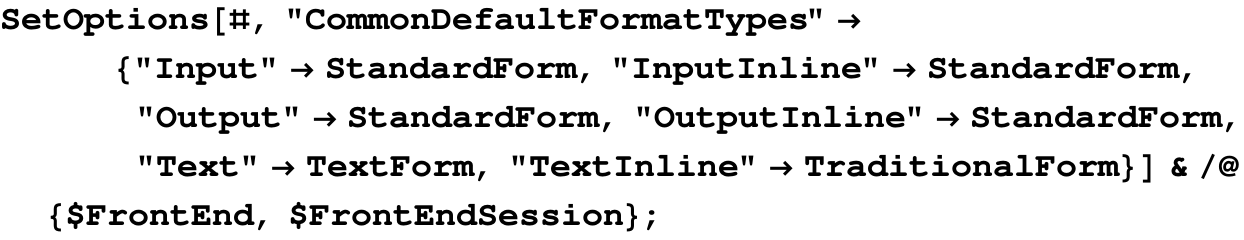}

\newpage
\subsection{Using \texttt{FeynCalcFormLink}}

\subsec{Examples 1}

\noindent Calculate a trace $\mbox{\tt
tr}(\gamma^\mu\gamma^\nu\gamma^\rho\gamma^\sigma\gamma^\tau\gamma^\sigma)$.
Using {\tt DiracTrace} in \textsc{FeynCalc} does not calculate immediately:
\vspace{0.5cm}

\vspace{.03cm}\noindent\includegraphics{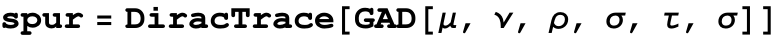}

\vspace{.03cm}\noindent\includegraphics{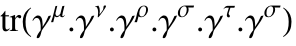}

\vspace{0.5cm}
\noindent Feeding this into {\tt FeynCalcFormLink} has enough information to tell {\sc FORM} to do the trace in {\tt D} dimensions:
\vspace{0.5cm}

\vspace{.03cm}\noindent\includegraphics{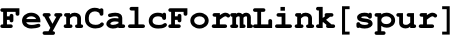}

\noindent\rule{\textwidth}{0.1pt}{\color{Code}
\texttt{\sf\\Symbol D;\\
Dimension D;\\
AutoDeclare Index lor;\\
Format Mathematica;\\
L resFL = (g$\_$(1,lor1)*g$\_$(1,lor2)*g$\_$(1,lor3)*g$\_$(1,lor4)*g$\_$(1,lor5)*g$\_$(1,lor4));\\
tracen,1;\\
contract 0;\\
.sort;\\
$\#$call put({``}$\%$E{''}, resFL)\\
$\#$fromexternal}}

\noindent\rule[.4\baselineskip]{\textwidth}{0.1pt}
\vspace{.03cm}\noindent\includegraphics{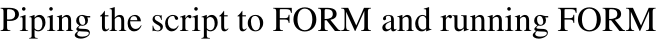}

\vspace{.03cm}\noindent\includegraphics{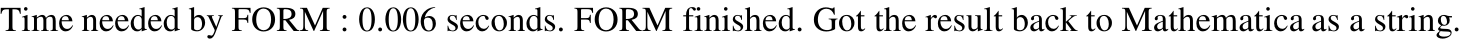}

\vspace{.03cm}\noindent\includegraphics{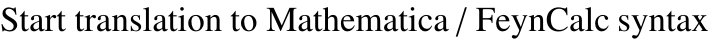}

\vspace{.03cm}\noindent\includegraphics{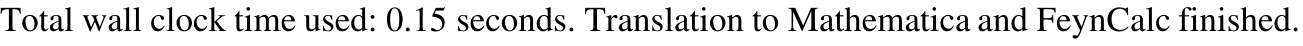}

\vspace{.03cm}\noindent\includegraphics{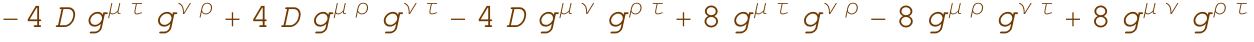}

\vspace{.03cm}\noindent\includegraphics{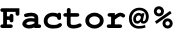}

\vspace{.03cm}\noindent\includegraphics{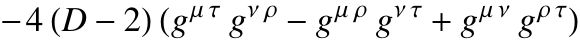}

\vspace{.03cm}\noindent\includegraphics{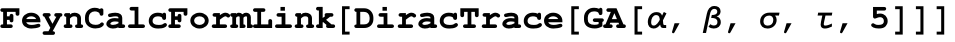}

\noindent\rule{\textwidth}{0.1pt}{\color{Code}
\texttt{\sf\\AutoDeclare Index lor;\\
Format Mathematica;\\
L resFL = (g$\_$(1,lor1)*g$\_$(1,lor2)*g$\_$(1,lor3)*g$\_$(1,lor4)*g5$\_$(1));\\
trace4,1;\\
contract 0;\\
.sort;\\
$\#$call put({``}$\%$E{''}, resFL)\\
$\#$fromexternal}}

\noindent\rule[.4\baselineskip]{\textwidth}{0.1pt}
\vspace{.03cm}\noindent\includegraphics{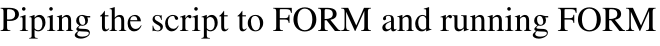}

\vspace{.03cm}\noindent\includegraphics{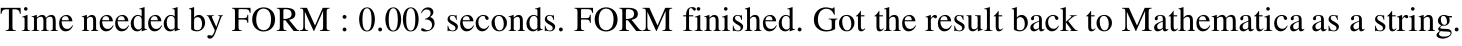}

\vspace{.03cm}\noindent\includegraphics{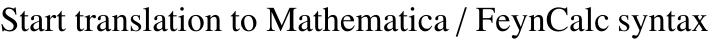}

\vspace{.03cm}\noindent\includegraphics{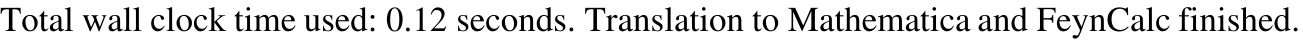}

\vspace{.03cm}\noindent\includegraphics{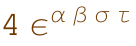}

\vspace{0.5cm}
\noindent Notice that loading {\sc FeynCalcFormLink} by default sets {\tt \$LeviCivitaSign=-I}, such that traces involving $\gamma^5$ agree.
\vspace{0.5cm}

\vspace{.03cm}\noindent\includegraphics{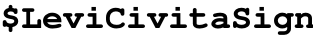}

\vspace{.03cm}\noindent\includegraphics{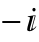}

\vspace{.03cm}\noindent\includegraphics{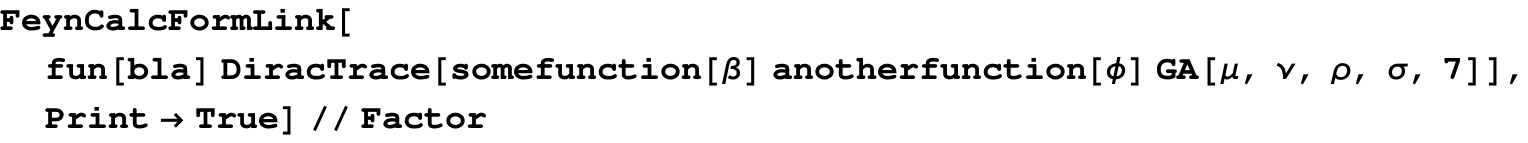}
\begin{samepage}
\noindent\rule{\textwidth}{0.1pt}{\color{Code}
\texttt{\sf\\Symbols bla,sym1,sym2;\\
AutoDeclare Index lor;\\
AutoDeclare Symbol sym;\\
CFunctions anotherfunction,fun,somefunction;\\
Format Mathematica;\\
L resFL = (anotherfunction(sym2)*fun(bla)*g$\_$(1,lor1)*g$\_$(1,lor2)*g$\_$(1,lor3)*g$\_$(1,lor4)\\
\hspace*{1.5cm}   *(g7$\_$(1)/2)*somefunction(sym1));\\
trace4,1;\\
contract 0;\\
.sort;\\
$\#$call put({``}$\%$E{''}, resFL)\\
$\#$fromexternal}}

\noindent\rule[.4\baselineskip]{\textwidth}{0.1pt}
\end{samepage}
\vspace{.03cm}\noindent\includegraphics{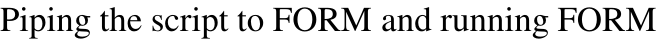}

\vspace{.03cm}\noindent\includegraphics{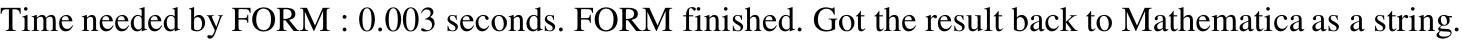}

\vspace{.03cm}\noindent\includegraphics{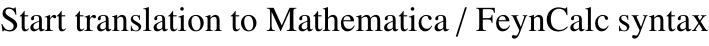}

\vspace{.03cm}\noindent\includegraphics{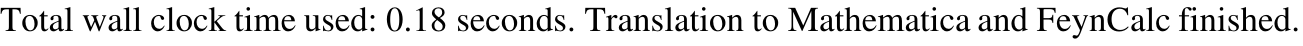}

\vspace{.03cm}\noindent\includegraphics{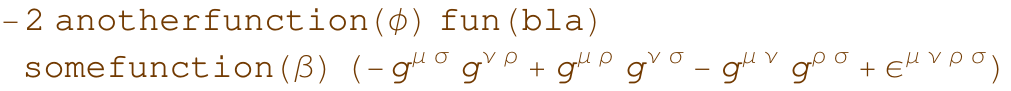}

\vspace{.03cm}\noindent\includegraphics{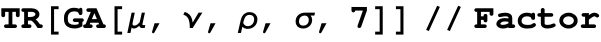}

\vspace{.03cm}\noindent\includegraphics{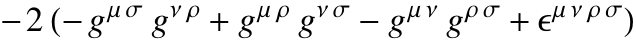}

\subsec{Example 2}

\vspace{.03cm}\noindent\includegraphics{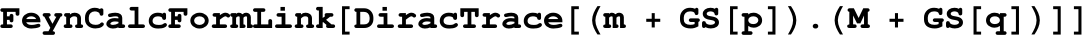}
\noindent\rule{\textwidth}{0.1pt}{\color{Code}
\texttt{\sf\\Symbols m,M;\\
Vectors p,q;\\
Format Mathematica;\\
L resFL = ((m*gi$\_$(1)+g$\_$(1,p))*(M*gi$\_$(1)+g$\_$(1,q)));\\
trace4,1;\\
contract 0;\\
.sort;\\
$\#$call put({``}$\%$E{''}, resFL)\\
$\#$fromexternal}}

\noindent\rule[.4\baselineskip]{\textwidth}{0.1pt}
\vspace{.03cm}\noindent\includegraphics{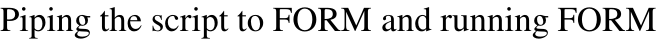}

\vspace{.03cm}\noindent\includegraphics{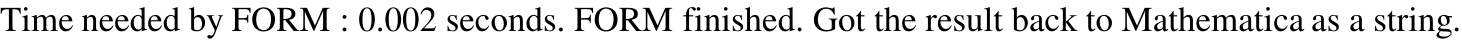}

\vspace{.03cm}\noindent\includegraphics{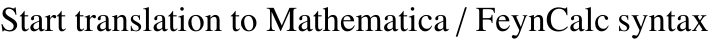}

\vspace{.03cm}\noindent\includegraphics{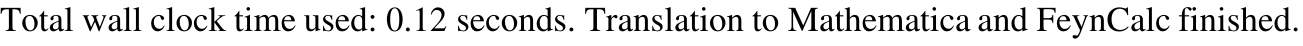}

\vspace{.03cm}\noindent\includegraphics{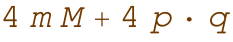}

\subsec{Example 3}

\noindent Define simple typesetting rules for p1, ... p8 :
\vspace{0.5cm}

\vspace{.03cm}\noindent\includegraphics{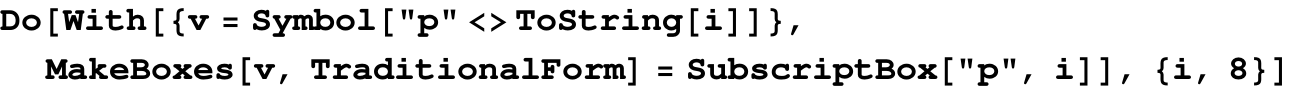}

\vspace{0.5cm}
\noindent Enter a trace $\displaystyle\mbox{tr}[(p\!\!\!/_1+m)\gamma^\mu(p\!\!\!/_2+k\!\!\!/+m)\gamma_\mu p\!\!\!/_2]$ like this in {\sc FeynCalc}:
\vspace{0.5cm}

\vspace{.03cm}\noindent\includegraphics{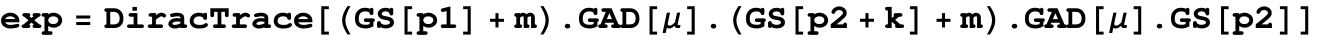}

\vspace{.03cm}\noindent\includegraphics{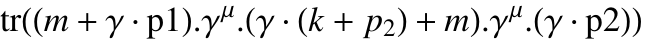}

\vspace{0.5cm}
\noindent Calculate it through {\sc FORM}:
\vspace{0.5cm}

\vspace{.03cm}\noindent\includegraphics{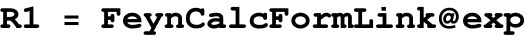}

\noindent\rule{\textwidth}{0.1pt}{\color{Code}
\texttt{\sf\\Symbols D,m;\\
Dimension D;\\
Vectors k,p1,p2;\\
AutoDeclare Index lor;\\
Format Mathematica;\\
L resFL = ((m*gi$\_$(1)+g$\_$(1,p1))*g$\_$(1,lor1)*(m*gi$\_$(1)+g$\_$(1,k)+g$\_$(1,p2))*g$\_$(1,lor1)\\
\hspace*{1.5cm} *g$\_$(1,p2));\\
tracen,1;\\
contract 0;\\
.sort;\\
$\#$call put({``}$\%$E{''}, resFL)\\
$\#$fromexternal}}

\vspace{0.5cm}
\noindent\rule[.4\baselineskip]{\textwidth}{0.1pt}
\vspace{.03cm}\noindent\includegraphics{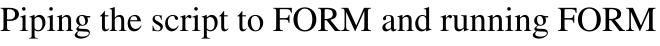}

\vspace{.03cm}\noindent\includegraphics{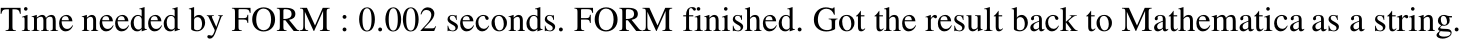}

\vspace{.03cm}\noindent\includegraphics{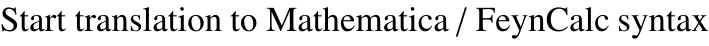}

\vspace{.03cm}\noindent\includegraphics{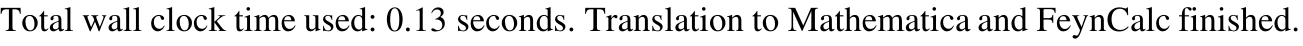}

\vspace{.03cm}\noindent\includegraphics{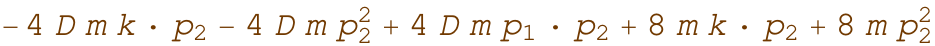}

\vspace{0.5cm}
\noindent Check with direct calculation in {\sc FeynCalc}:
\vspace{0.5cm}

\vspace{.03cm}\noindent\includegraphics{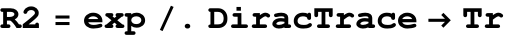}

\vspace{.03cm}\noindent\includegraphics{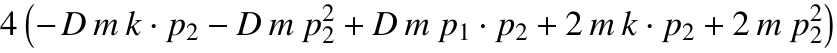}

\vspace{.03cm}\noindent\includegraphics{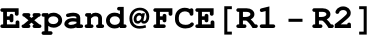}

\vspace{.03cm}\noindent\includegraphics{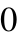}

\vspace{.03cm}\noindent\includegraphics{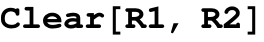}

\subsection{Using \texttt{RunForm}}

\noindent
{\tt RunForm} runs a {\sc FORM} program without \textit{MathLink}, but by
calling {\sc FORM} directly.
\vspace{.2cm}

\vspace{.03cm}\noindent\includegraphics{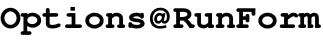}

\vspace{.03cm}\noindent\includegraphics{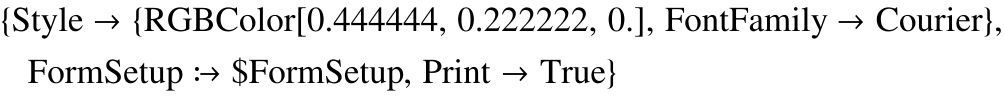}

\vspace{.03cm}\noindent\includegraphics{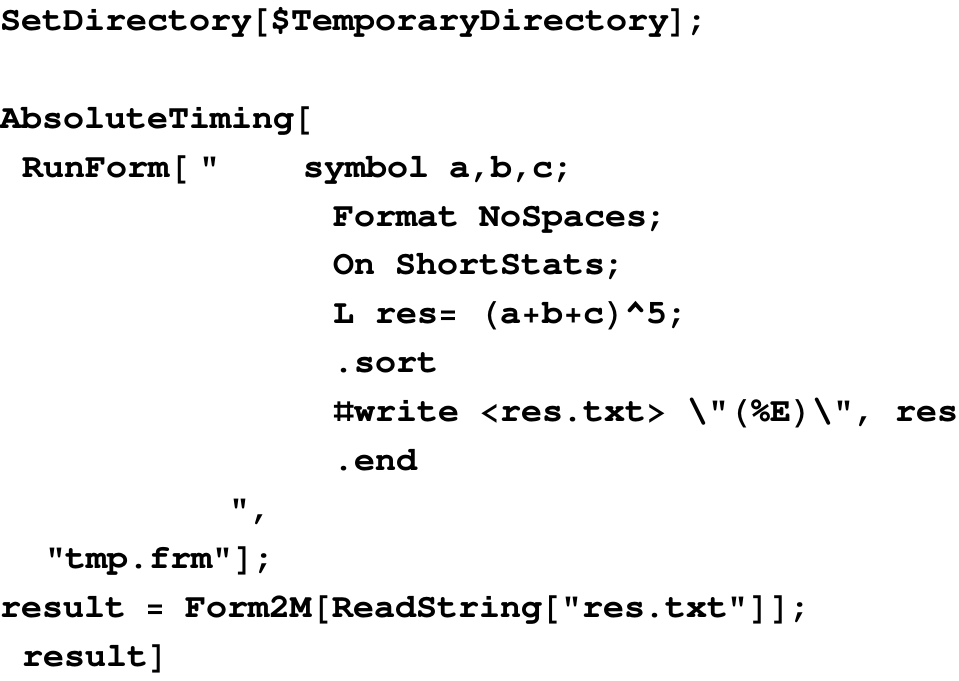}

\noindent\rule[.4\baselineskip]{\textwidth}{0.1pt}\\{\color{Code}
\texttt{\sf{}FORM 4.0 (Aug 31 2012) 32-bits \hfill Run: Wed Dec 12 13:21:06 2012\\
\hspace*{5.ex} symbol a,b,c;\\
\hspace*{5.ex} Format NoSpaces;\\
\hspace*{5.ex} On ShortStats;\\
\hspace*{5.ex} L res= (a+b+c)${}^{\wedge}$5;\\
\hspace*{5.ex} .sort\\
\hspace*{5.ex} 0.00s \hfill 1$>$ \hfill 21-{}-$>$ \hfill 21: \hfill 380 res \\
\hspace*{5.ex} $\#$write $<$res.txt$>$ {``}($\%$E){''}, res\\
\hspace*{5.ex} .end\\
\hspace*{5.ex} 0.00s \hfill 21$>$ \hfill 21-{}-$>$ \hfill 21: \hfill 380 res \\
\hspace*{1.ex} 0.00 sec out of 0.00 sec}}

\noindent\rule[.4\baselineskip]{\textwidth}{0.1pt}

\vspace{.03cm}\noindent\includegraphics{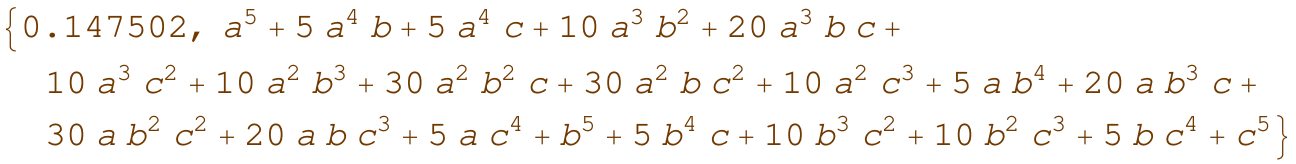}

\vspace{.03cm}\noindent\includegraphics{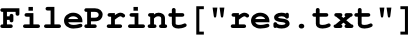}

\vspace{.03cm}\noindent\includegraphics{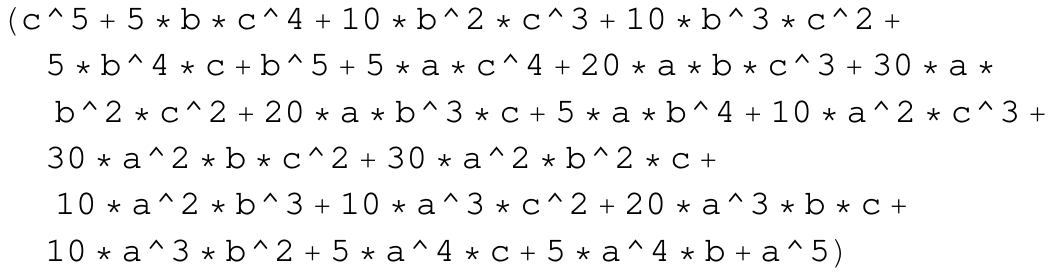}

\vspace{.03cm}\noindent\includegraphics{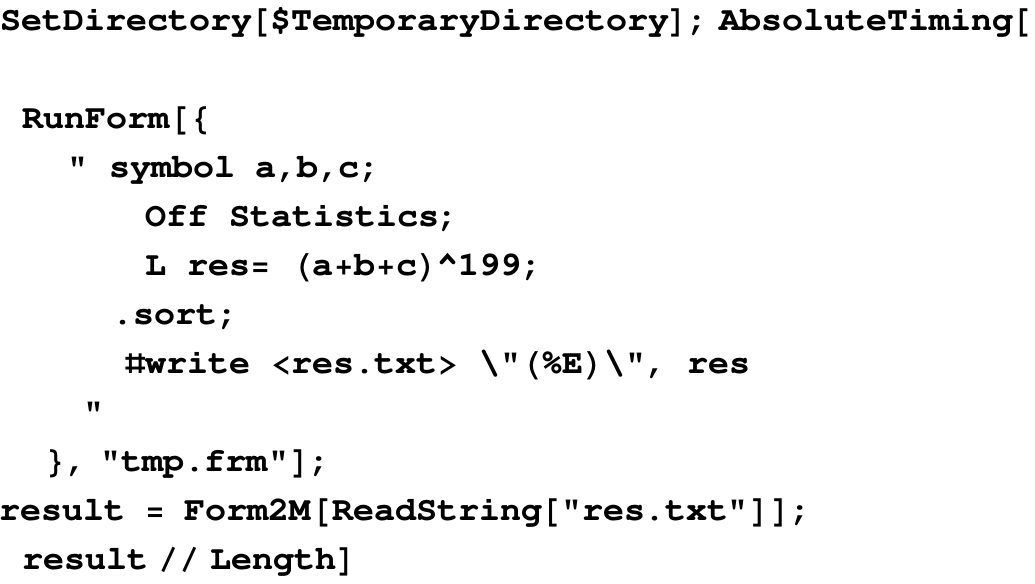}

\noindent\rule{\textwidth}{0.1pt}\\{\color{Code}
\texttt{\sf{}FORM 4.0 (Aug 31 2012) 32-bits \hfill Run: Wed Dec 12 13:21:06 2012\\
\hspace*{2.ex} symbol a,b,c;\\
\hspace*{2.ex} Off Statistics;\\
\hspace*{2.ex} L res= (a+b+c)${}^{\wedge}$199;\\
\hspace*{2.ex} .sort;\\
\hspace*{2.ex} $\#$write $<$res.txt$>$ {``}($\%$E){''}, res\\
\hspace*{2.ex} .end\\
\hspace*{1.ex} 1.21 sec out of 1.22 sec}}

\noindent\rule[.4\baselineskip]{\textwidth}{0.1pt}

\vspace{.03cm}\noindent\includegraphics{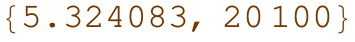}

\vspace{.03cm}\noindent\includegraphics{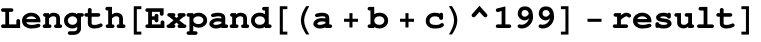}

\vspace{.03cm}\noindent\includegraphics{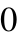}

\vspace{.03cm}\noindent\includegraphics{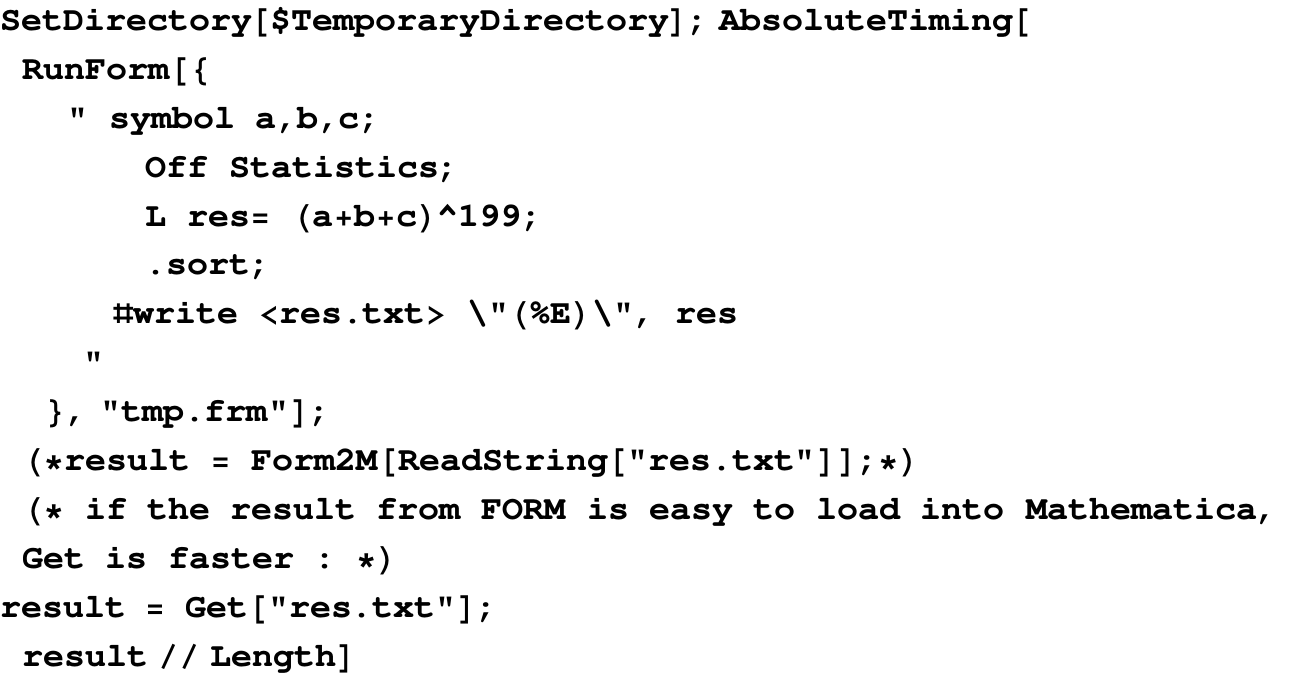}

\noindent\rule{\textwidth}{0.1pt}\\{\color{Code}
\texttt{\sf{}FORM 4.0 (Aug 31 2012) 32-bits \hfill Run: Wed Dec 12 13:21:12 2012\\
\hspace*{2.ex} symbol a,b,c;\\
\hspace*{2.ex} Off Statistics;\\
\hspace*{2.ex} L res= (a+b+c)${}^{\wedge}$199;\\
\hspace*{2.ex} .sort;\\
\hspace*{2.ex} $\#$write $<$res.txt$>$ {``}($\%$E){''}, res\\
\hspace*{2.ex} .end\\
\hspace*{1.ex} 1.21 sec out of 1.21 sec}}

\noindent\rule[.4\baselineskip]{\textwidth}{0.1pt}

\vspace{.03cm}\noindent\includegraphics{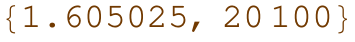}

\vspace{.03cm}\noindent\includegraphics{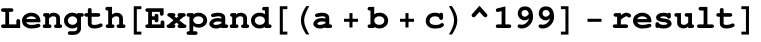}

\vspace{.03cm}\noindent\includegraphics{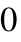}

\noindent For such simple algebraic operations \textit{Mathematica} is actually much faster.

\vspace{.03cm}\noindent\includegraphics{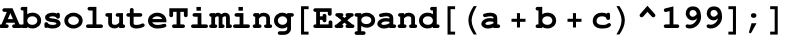}

\vspace{.03cm}\noindent\includegraphics{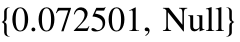}

\vspace{.03cm}\noindent\includegraphics{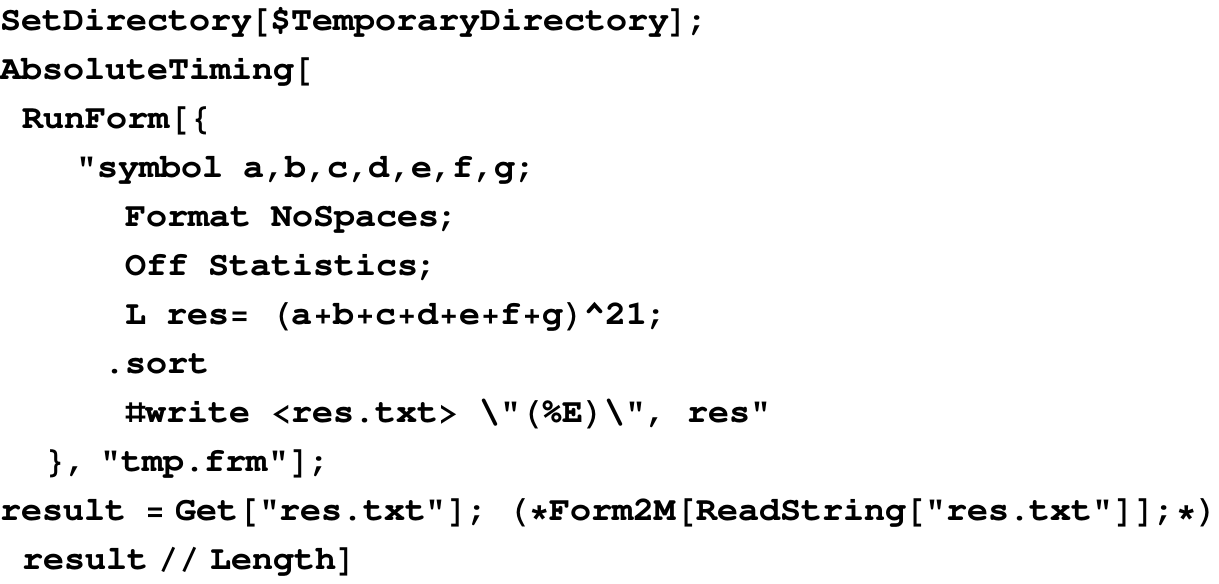}

\noindent\rule{\textwidth}{0.1pt}\\{\color{Code}
\texttt{\sf{}FORM 4.0 (Aug 31 2012) 32-bits \hfill Run: Wed Dec 12 13:21:14 2012\\
\hspace*{2.ex} symbol a,b,c,d,e,f,g;\\
\hspace*{2.ex} Format NoSpaces;\\
\hspace*{2.ex} Off Statistics;\\
\hspace*{2.ex} L res= (a+b+c+d+e+f+g)${}^{\wedge}$21;\\
\hspace*{2.ex} .sort\\
\hspace*{5.ex} $\#$write $<$res.txt$>$ {``}($\%$E){''}, res\\
\hspace*{2.ex} .end\\
\hspace*{1.ex} 1.34 sec out of 1.35 sec}}

\noindent\rule[.4\baselineskip]{\textwidth}{0.1pt}

\vspace{.03cm}\noindent\includegraphics{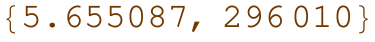}

\vspace{.03cm}\noindent\includegraphics{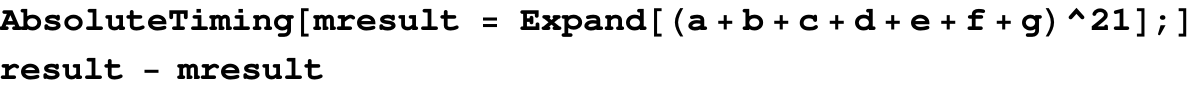}

\vspace{.03cm}\noindent\includegraphics{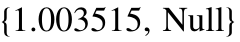}

\vspace{.03cm}\noindent\includegraphics{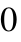}

\section{Application to Tree Level Processes}
\subsection{$e^+e^-\to\tau^+\tau^-\to u\,\bar{d}\,\mu\,\bar{\nu}_\mu\,\nu_\tau\,\bar{\nu}_\tau$}
\noindent Let us consider some applications of \textsc{FormLink} combined with \textsc{FeynCalc}. We take the process: $e^+e^-\to\tau^+\tau^-\to u\,\bar{d}\,\mu\,\bar{\nu}_\mu\,\nu_\tau\,\bar{\nu}_\tau$ as an example, which has been considered in the \textsc{Form} courses\cite{FormCourse}. We can express the squared amplitude as:
\vspace{-.0cm}
\def\p{p\!\!\!\slash}
\def\q{q\!\!\!\slash}
\def\tr{\mbox{Tr}}
{\small\begin{eqnarray}
&&\frac{1}{2^{16}}\;\tr\big[(\p_2-m_e) \gamma^{\mu_1}(\p_1+m_e) \gamma^{\nu_1} \big]\nonumber\\
&*& \tr\big[ (\p_3+m_3) \gamma^{\mu_2} \gamma_7 (\q_1+m_\tau) \gamma^{\mu_1} (-\q_2+m_\tau) \gamma^{\mu_3} \gamma_7 (\p_6-m_6) \gamma^{\nu_3} \gamma^7 (-\q_2+m_\tau) \gamma^{\nu_1} (\q_1+m_\tau) \gamma^{\nu_2} \gamma_7 \big] \nonumber\\
&*& \tr\big[(\p_4+m_4) \gamma^{\mu_2} \gamma_7 (\p_5-m_5) \gamma^{\nu_2} \gamma_7\big] \nonumber\\
&*& \tr\big[(\p_7+m_7) \gamma^{\mu_3} \gamma_7 (\p_8-m_8) \gamma^{\nu_3} \gamma_7\big] \label{Squared Amplitude}
\end{eqnarray}}
\noindent The expression in (\ref{Squared Amplitude}) can be easily translated to \textsc{FeynCalc} syntax and then executed by \texttt{FeynCalcFormLink},
it takes a fraction of a second to run the code. The program is:
\vspace{0.5cm }

\vspace{.03cm}\noindent\includegraphics{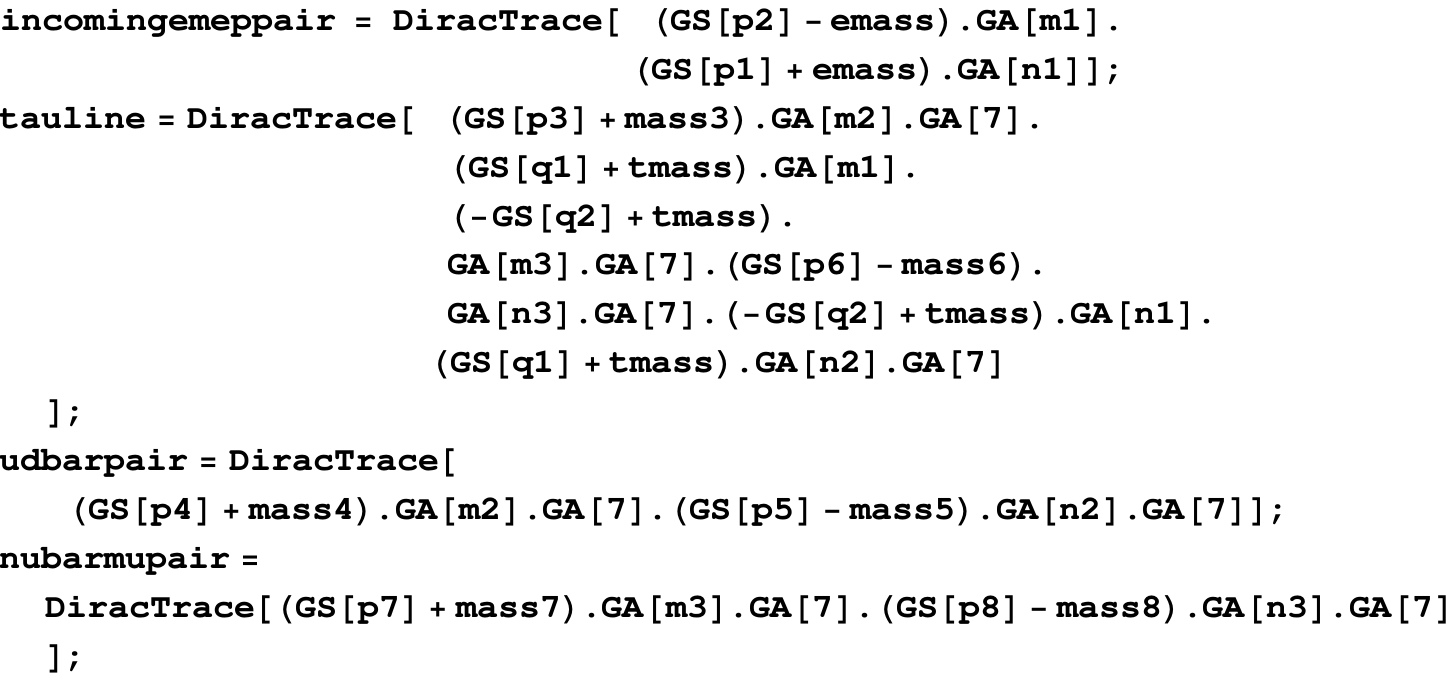}

\vspace{.03cm}\noindent\includegraphics{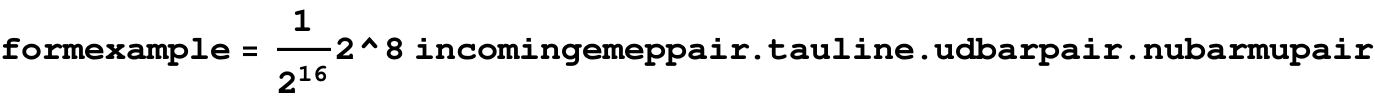}

\vspace{.03cm}\noindent\includegraphics{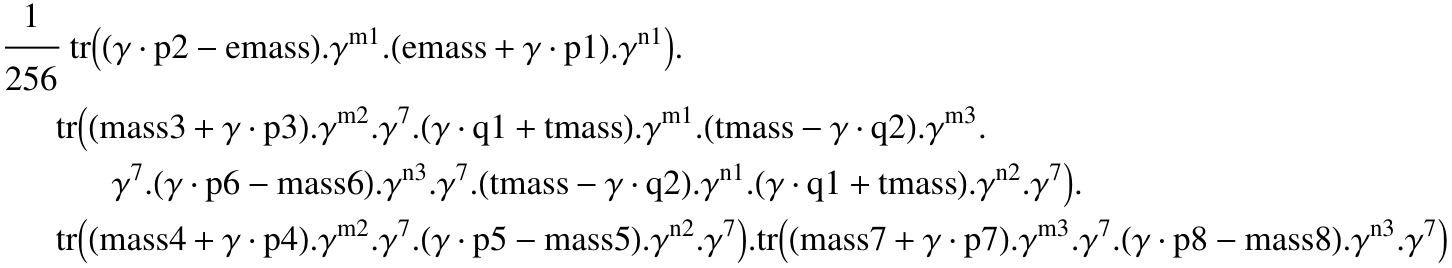}

\vspace{.03cm}\noindent\includegraphics{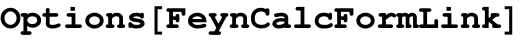}

\vspace{.03cm}\noindent\includegraphics{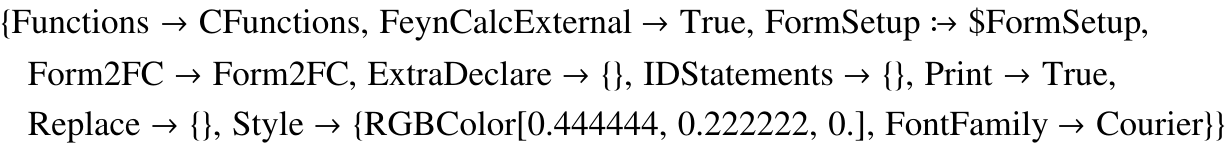}

\vspace{.03cm}\noindent\includegraphics{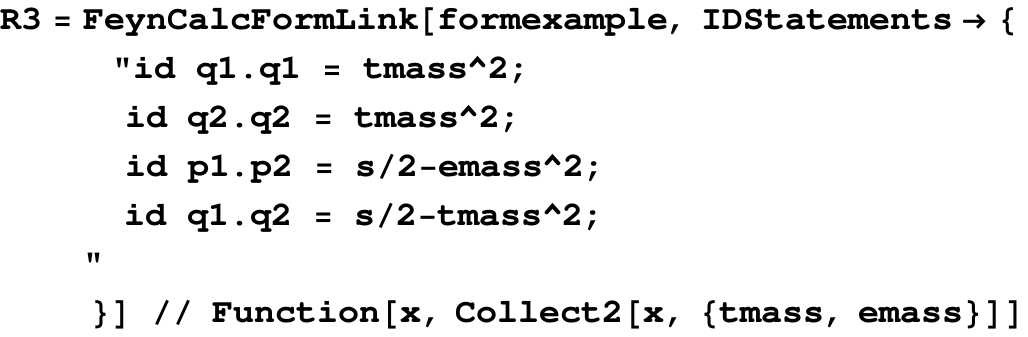}

\noindent\rule{\textwidth}{0.1pt}{\color{Code}
\texttt{\sf\\Symbols emass,mass3,mass4,mass5,mass6,mass7,mass8,s,tmass;\\
Indices m1,m2,m3,n1,n2,n3;\\
Vectors p1,p2,p3,p4,p5,p6,p7,p8,q1,q2;\\
Format Mathematica;\\
L resFL = (((-(emass*gi$\_$(1))+g$\_$(1,p2))*g$\_$(1,m1)*(emass*gi$\_$(1)+g$\_$(1,p1))\\
\hspace*{1.5cm}	*g$\_$(1,n1)*(mass4*gi$\_$(2)+g$\_$(2,p4))*g$\_$(2,m2)*(g7$\_$(2)/2)*(-(mass5*gi$\_$(2))\\
\hspace*{1.5cm}	+g$\_$(2,p5))*g$\_$(2,n2)*(g7$\_$(2)/2)*(mass7*gi$\_$(3)+g$\_$(3,p7))*g$\_$(3,m3)\\
\hspace*{1.5cm}	*(g7$\_$(3)/2)*(-(mass8*gi$\_$(3))+g$\_$(3,p8))*g$\_$(3,n3)*(g7$\_$(3)/2)*(mass3\\
\hspace*{1.5cm}	*gi$\_$(4)+g$\_$(4,p3))*g$\_$(4,m2)*(g7$\_$(4)/2)*(tmass*gi$\_$(4)+g$\_$(4,q1))\\
\hspace*{1.5cm}	*g$\_$(4,m1)*(tmass*gi$\_$(4)-g$\_$(4,q2))*g$\_$(4,m3)*(g7$\_$(4)/2)*(-(mass6\\
\hspace*{1.5cm}	*gi$\_$(4))+g$\_$(4,p6))*g$\_$(4,n3)*(g7$\_$(4)/2)*(tmass*gi$\_$(4)-g$\_$(4,q2))\\
\hspace*{1.5cm}	*g$\_$(4,n1)*(tmass*gi$\_$(4)+g$\_$(4,q1))*g$\_$(4,n2)*(g7$\_$(4)/2))/256);\\
trace4,1;\\
trace4,2;\\
trace4,3;\\
trace4,4;\\
contract 0;\\
.sort;\\
id q1.q1 = tmass${}^{\wedge}$2;\\
id q2.q2 = tmass${}^{\wedge}$2;\\
id p1.p2 = s/2-emass${}^{\wedge}$2;\\
id q1.q2 = s/2-tmass${}^{\wedge}$2;\\
.sort;\\
$\#$call put({``}$\%$E{''}, resFL)\\
$\#$fromexternal}}

\noindent\rule[.4\baselineskip]{\textwidth}{0.1pt}
\vspace{.03cm}\noindent\includegraphics{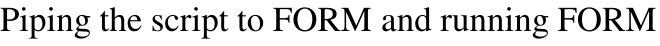}

\vspace{.03cm}\noindent\includegraphics{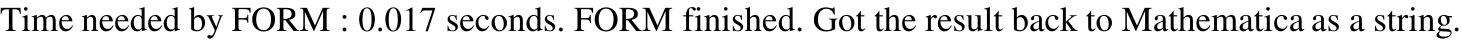}

\vspace{.03cm}\noindent\includegraphics{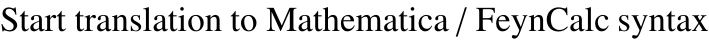}

\vspace{.03cm}\noindent\includegraphics{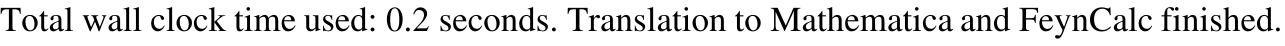}

\vspace{.03cm}\noindent\includegraphics{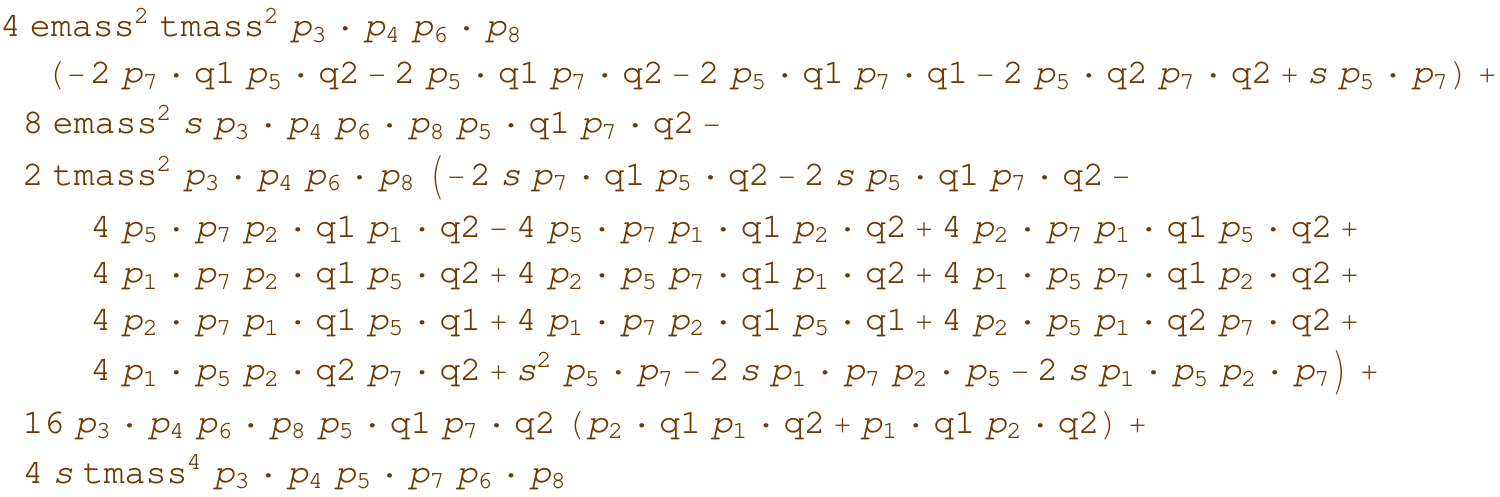}

\vspace{.03cm}\noindent\includegraphics{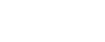}

\vspace{.03cm}\noindent\includegraphics{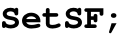}

\vspace{.03cm}\noindent\includegraphics{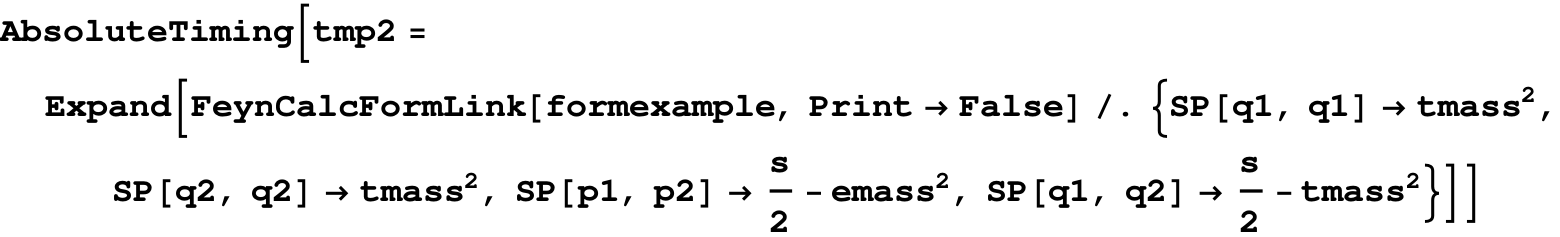}

\vspace{.03cm}\noindent\includegraphics{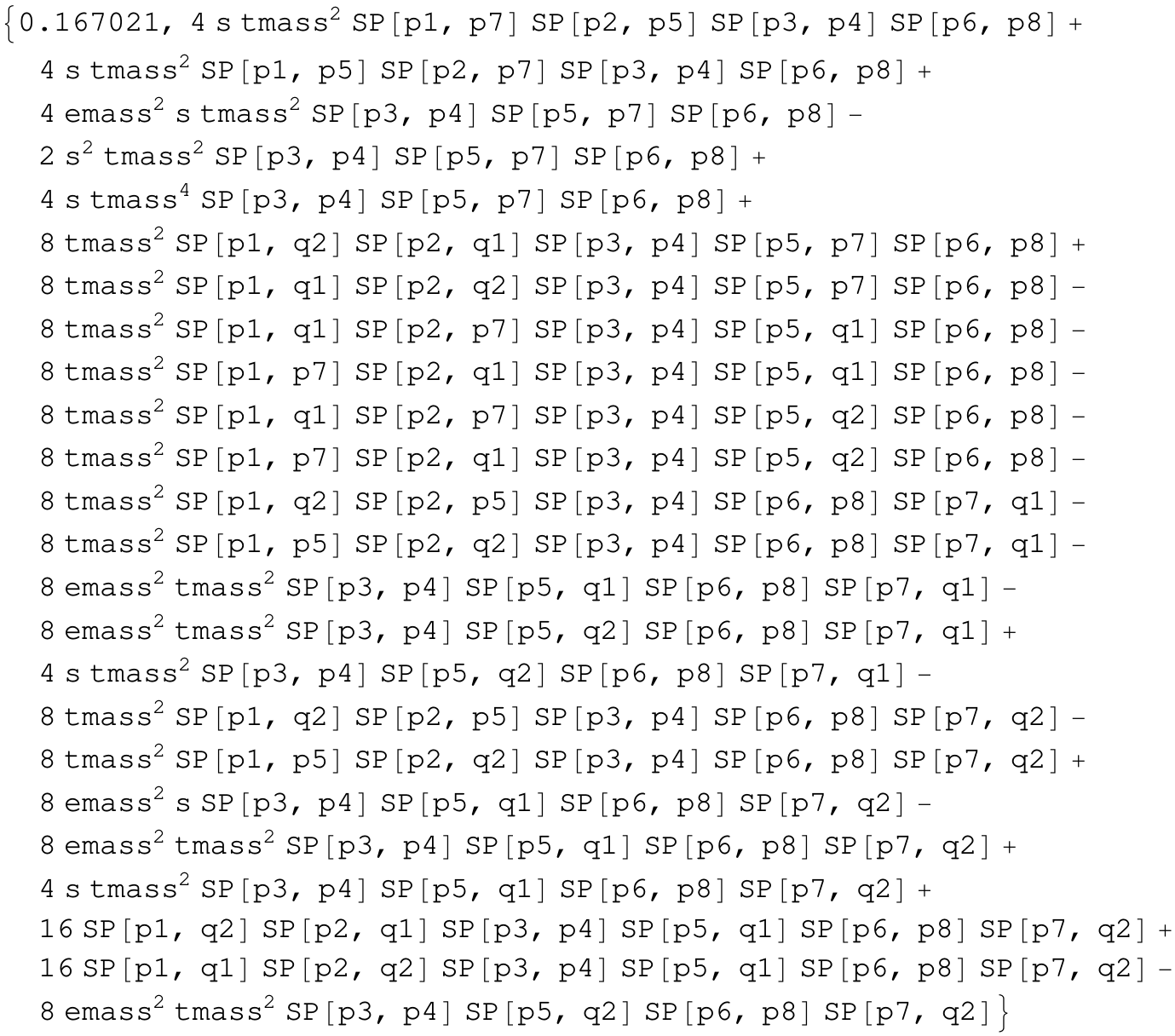}

\vspace{.03cm}\noindent\includegraphics{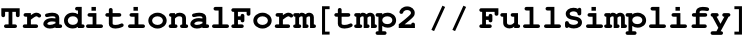}

\vspace{.03cm}\noindent\includegraphics{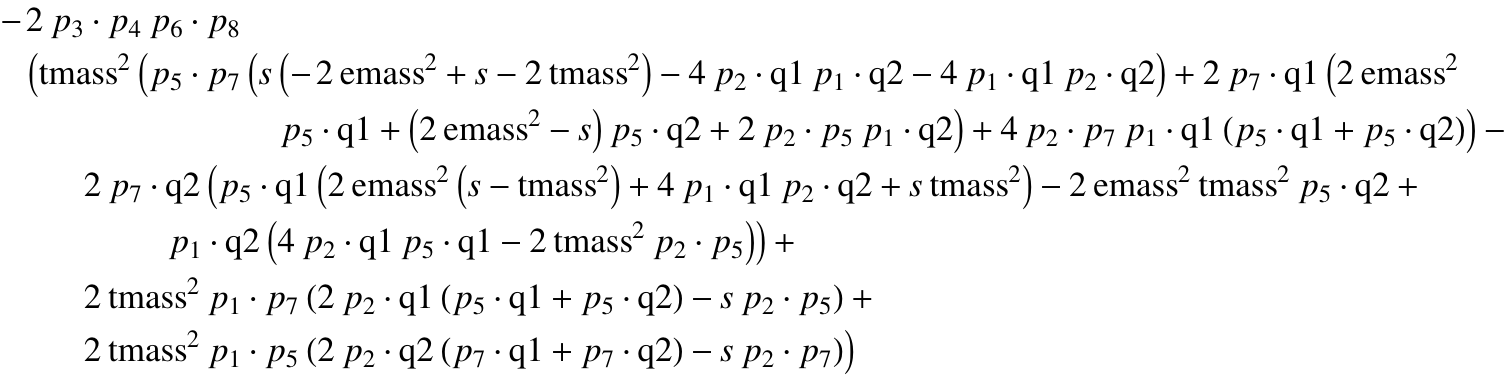}

\vspace{.03cm}\noindent\includegraphics{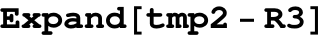}

\vspace{.03cm}\noindent\includegraphics{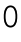}

\subsection{Double Bremsstrahlung in $\tau$ leptonic radiative decay}
\noindent This example is provided by Matteo Fael, it refers to a part of the calculation done in \cite{Fischer:1994pn}.

\subsec{Initialization}

\vspace{.03cm}\noindent\includegraphics{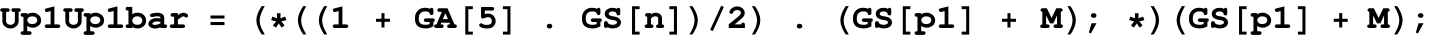}

\vspace{.03cm}\noindent\includegraphics{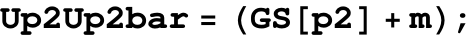}

\subsec{Bremsstrahlung}

\subsec{Amplitudes without spinors}

\vspace{.03cm}\noindent\includegraphics{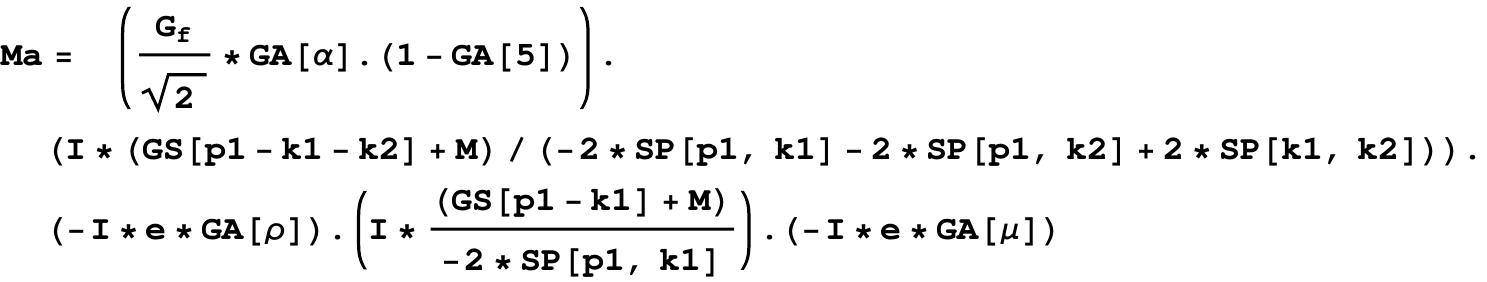}

\vspace{.03cm}\noindent\includegraphics{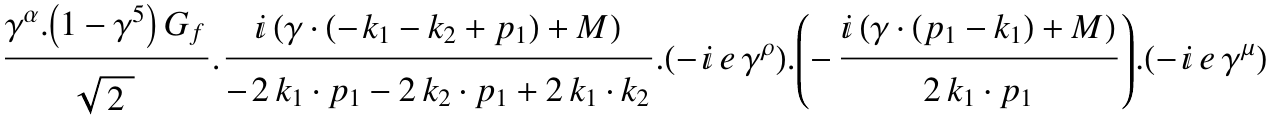}

\vspace{.03cm}\noindent\includegraphics{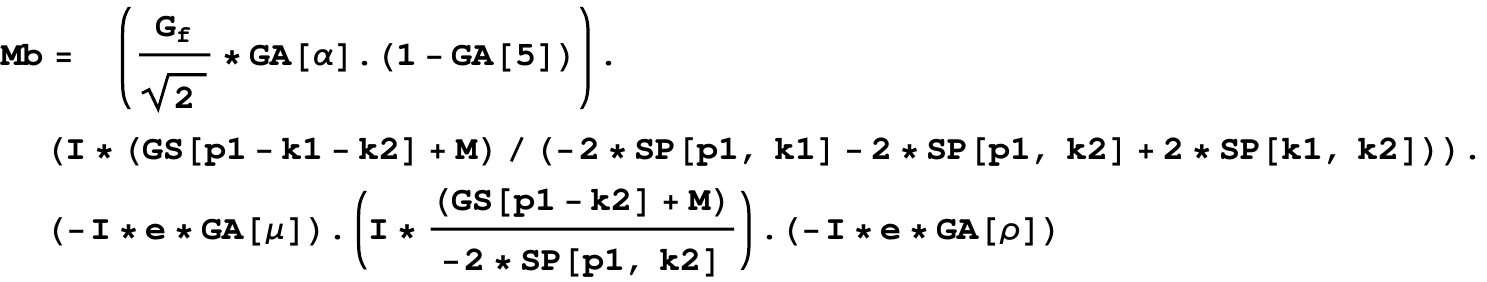}

\vspace{.03cm}\noindent\includegraphics{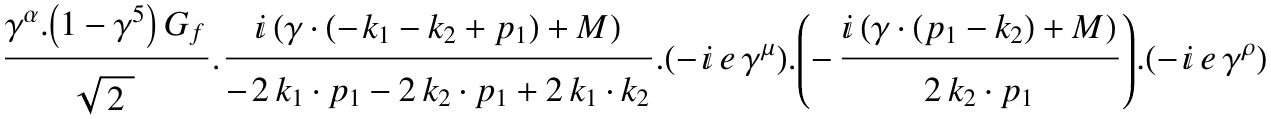}

\vspace{.03cm}\noindent\includegraphics{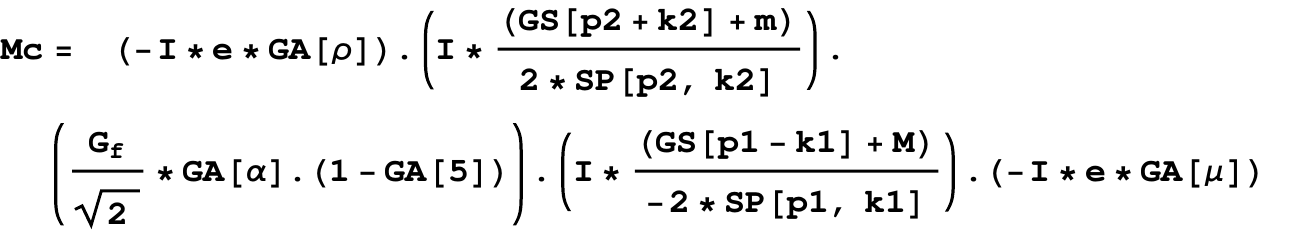}

\vspace{.03cm}\noindent\includegraphics{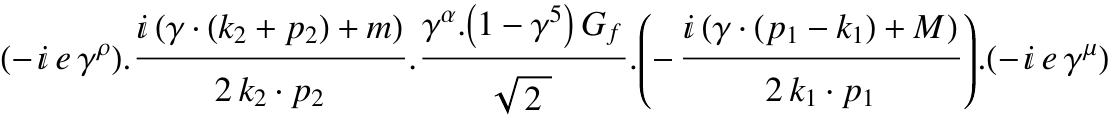}

\vspace{.03cm}\noindent\includegraphics{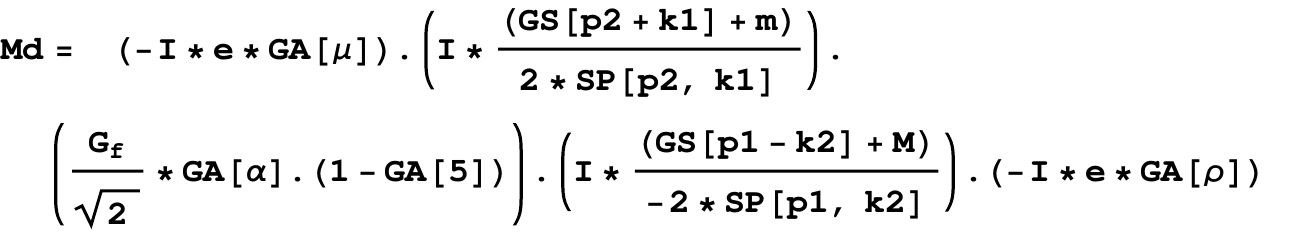}

\vspace{.03cm}\noindent\includegraphics{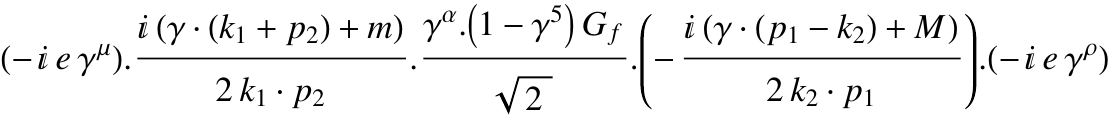}

\vspace{.03cm}\noindent\includegraphics{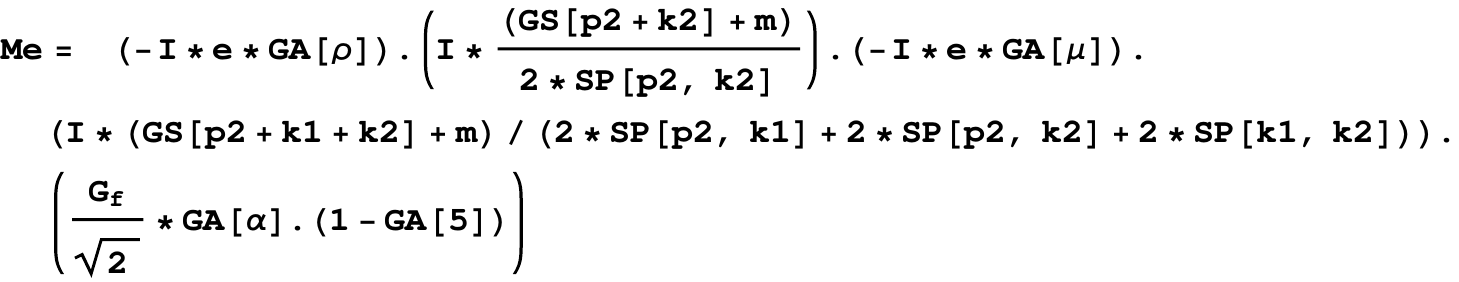}

\vspace{.03cm}\noindent\includegraphics{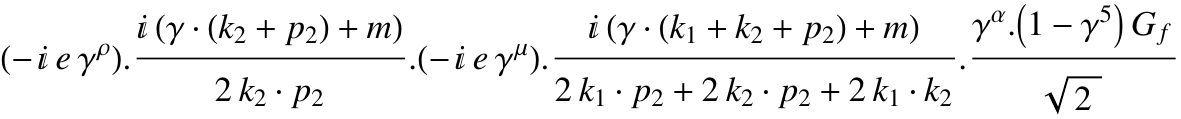}

\vspace{.03cm}\noindent\includegraphics{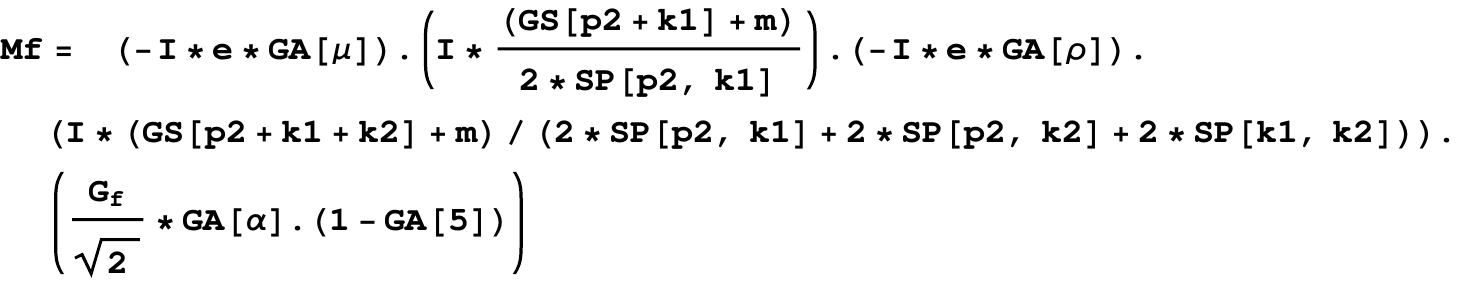}

\vspace{.03cm}\noindent\includegraphics{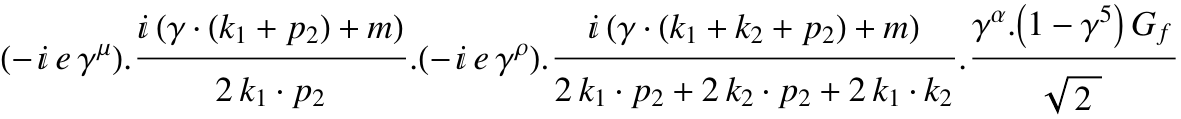}

\vspace{.03cm}\noindent\includegraphics{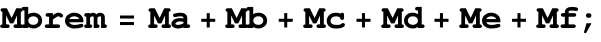}

\vspace{.03cm}\noindent\includegraphics{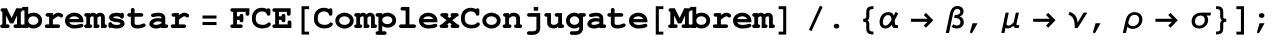}

\vspace{.03cm}\noindent\includegraphics{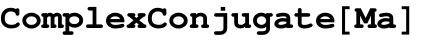}

\vspace{.03cm}\noindent\includegraphics{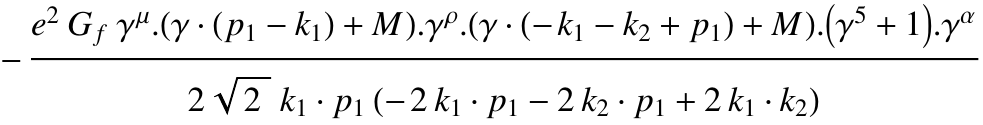}

\subsec{Squared Amplitude}

\noindent ATTENTION! The tensor of Neutrini changes in the double photon emission!

\vspace{.03cm}\noindent\includegraphics{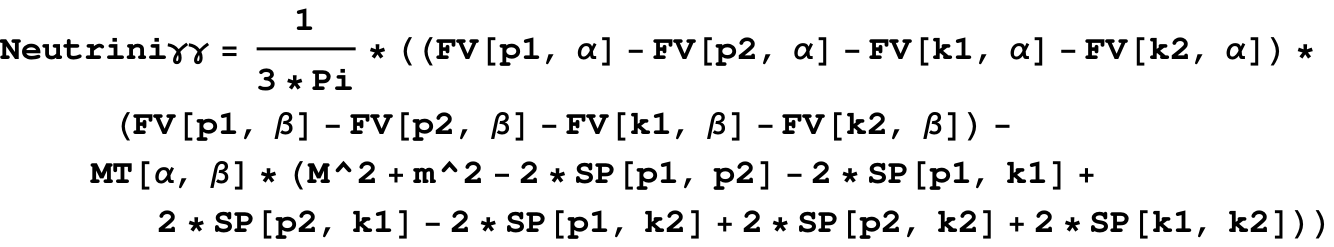}

\vspace{.03cm}\noindent\includegraphics{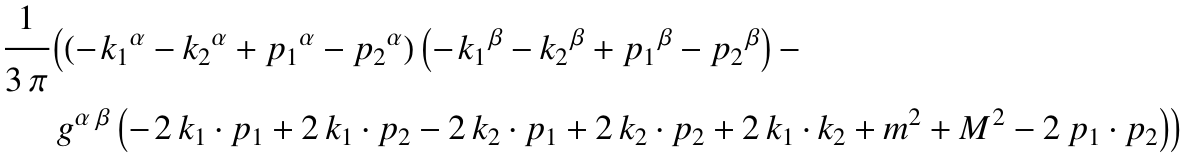}

\vspace{.03cm}\noindent\includegraphics{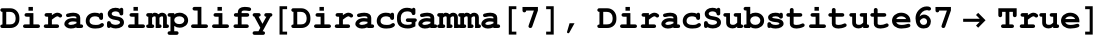}

\vspace{.03cm}\noindent\includegraphics{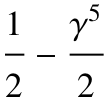}

\vspace{.03cm}\noindent\includegraphics{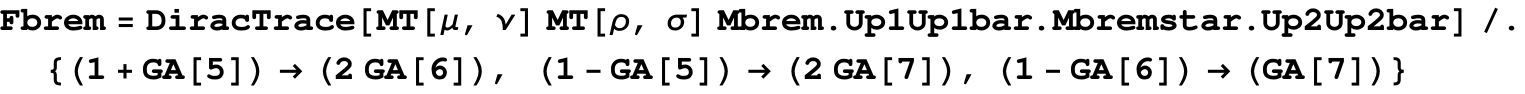}

\vspace{.03cm}\noindent\includegraphics{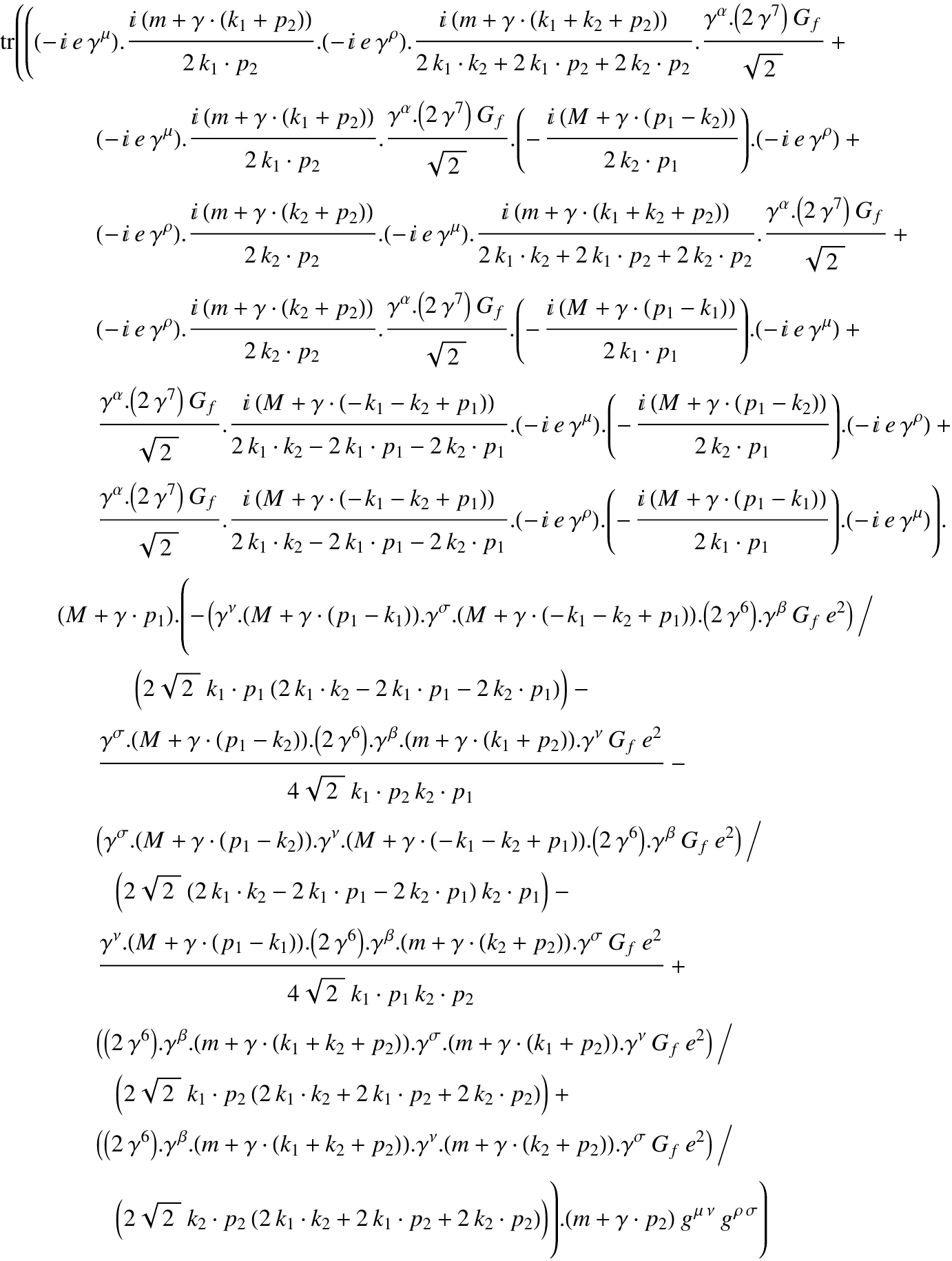}

\vspace{.03cm}\noindent\includegraphics{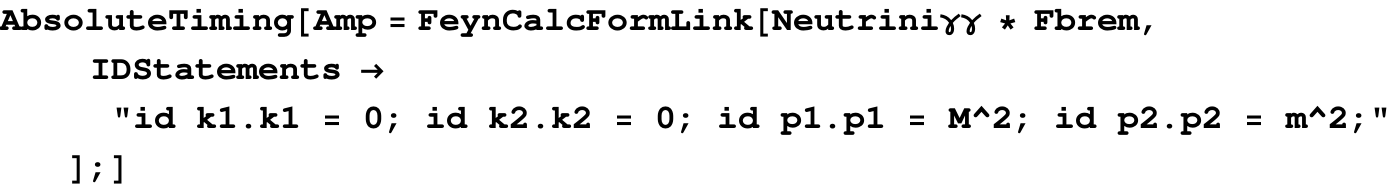}

\begin{samepage}
\noindent\rule{\textwidth}{0.1pt}{\color{Code}
\texttt{\sf\\Symbols e,Gsubf,m,M;\\
Vectors k1,k2,p1,p2;\\
AutoDeclare Index lor;\\
Format Mathematica;\\
L resFL = ((d$\_$(lor3,lor4)*d$\_$(lor5,lor6)*(-((e${}^{\wedge}$2*Gsubf*g$\_$(1,lor1)*(g7$\_$(1)/2)*(M-g$\_$(1,k1)\\
\hspace*{1.cm}	-g$\_$(1,k2)+g$\_$(1,p1))*g$\_$(1,lor5)*(M-g$\_$(1,k1)+g$\_$(1,p1))*g$\_$(1,lor3))/(sqrt$\_$(2)\\
\hspace*{1.cm}	*p1(k1)*(2*k2(k1)-2*p1(k1)-2*p1(k2))))-(e${}^{\wedge}$2*Gsubf*g$\_$(1,lor1)*(g7$\_$(1)/2)\\
\hspace*{1.cm}	*(M-g$\_$(1,k1)-g$\_$(1,k2)+g$\_$(1,p1))*g$\_$(1,lor3)*(M-g$\_$(1,k2)+g$\_$(1,p1))\\
\hspace*{1.cm}	*g$\_$(1,lor5))/(sqrt$\_$(2)*(2*k2(k1)-2*p1(k1)-2*p1(k2))*p1(k2))-(e${}^{\wedge}$2*Gsubf\\
\hspace*{1.cm}	*g$\_$(1,lor3)*(m+g$\_$(1,k1)+g$\_$(1,p2))*g$\_$(1,lor1)*(g7$\_$(1)/2)*(M-g$\_$(1,k2)\\
\hspace*{1.cm}	+g$\_$(1,p1))*g$\_$(1,lor5))/(2*sqrt$\_$(2)*p1(k2)*p2(k1))-(e${}^{\wedge}$2*Gsubf*g$\_$(1,lor5)\\
\hspace*{1.cm}	*(m+g$\_$(1,k2)+g$\_$(1,p2))*g$\_$(1,lor1)*(g7$\_$(1)/2)*(M-g$\_$(1,k1)+g$\_$(1,p1))\\
\hspace*{1.cm}	*g$\_$(1,lor3))/(2*sqrt$\_$(2)*p1(k1)*p2(k2))+(e${}^{\wedge}$2*Gsubf*g$\_$(1,lor3)*(m+g$\_$(1,k1)\\
\hspace*{1.cm}	+g$\_$(1,p2))*g$\_$(1,lor5)*(m+g$\_$(1,k1)+g$\_$(1,k2)+g$\_$(1,p2))*g$\_$(1,lor1)\\
\hspace*{1.cm}	*(g7$\_$(1)/2))/(sqrt$\_$(2)*p2(k1)*(2*k2(k1)+2*p2(k1)+2*p2(k2)))+(e${}^{\wedge}$2*Gsubf\\
\hspace*{1.cm}	*g$\_$(1,lor5)*(m+g$\_$(1,k2)+g$\_$(1,p2))*g$\_$(1,lor3)*(m+g$\_$(1,k1)+g$\_$(1,k2)+g$\_$(1,p2))\\
\hspace*{1.cm}	*g$\_$(1,lor1)*(g7$\_$(1)/2))/(sqrt$\_$(2)*p2(k2)*(2*k2(k1)+2*p2(k1)+2*p2(k2))))\\
\hspace*{1.cm}	*(M*gi$\_$(1)+g$\_$(1,p1))*(-((e${}^{\wedge}$2*Gsubf*g$\_$(1,lor4)*(M-g$\_$(1,k1)+g$\_$(1,p1))\\
\hspace*{1.cm}	*g$\_$(1,lor6)*(M-g$\_$(1,k1)-g$\_$(1,k2)+g$\_$(1,p1))*(g6$\_$(1)/2)*g$\_$(1,lor2))/(sqrt$\_$(2)\\
\hspace*{1.cm}	*p1(k1)*(2*k2(k1)-2*p1(k1)-2*p1(k2))))-(e${}^{\wedge}$2*Gsubf*g$\_$(1,lor6)*(M-g$\_$(1,k2)\\
\hspace*{1.cm}	+g$\_$(1,p1))*g$\_$(1,lor4)*(M-g$\_$(1,k1)-g$\_$(1,k2)+g$\_$(1,p1))*(g6$\_$(1)/2)\\
\hspace*{1.cm}	*g$\_$(1,lor2))/(sqrt$\_$(2)*(2*k2(k1)-2*p1(k1)-2*p1(k2))*p1(k2))-(e${}^{\wedge}$2*Gsubf*g$\_$(1,lor6)\\
\hspace*{1.cm}	*(M-g$\_$(1,k2)+g$\_$(1,p1))*(g6$\_$(1)/2)*g$\_$(1,lor2)*(m+g$\_$(1,k1)+g$\_$(1,p2))\\
\hspace*{1.cm}	*g$\_$(1,lor4))/(2*sqrt$\_$(2)*p1(k2)*p2(k1))-(e${}^{\wedge}$2*Gsubf*g$\_$(1,lor4)*(M-g$\_$(1,k1)\\
\hspace*{1.cm}	+g$\_$(1,p1))*(g6$\_$(1)/2)*g$\_$(1,lor2)*(m+g$\_$(1,k2)+g$\_$(1,p2))*g$\_$(1,lor6))/(2\\
\hspace*{1.cm}	*sqrt$\_$(2)*p1(k1)*p2(k2))+(e${}^{\wedge}$2*Gsubf*(g6$\_$(1)/2)*g$\_$(1,lor2)\\
\hspace*{1.cm}	*(m+g$\_$(1,k1)+g$\_$(1,k2)+g$\_$(1,p2))*g$\_$(1,lor6)*(m+g$\_$(1,k1)+g$\_$(1,p2))\\
\hspace*{1.cm}	*g$\_$(1,lor4))/(sqrt$\_$(2)*p2(k1)*(2*k2(k1)+2*p2(k1)+2*p2(k2)))+(e${}^{\wedge}$2*Gsubf\\
\hspace*{1.cm}	*(g6$\_$(1)/2)*g$\_$(1,lor2)*(m+g$\_$(1,k1)+g$\_$(1,k2)+g$\_$(1,p2))*g$\_$(1,lor4)\\
\hspace*{1.cm}	*(m+g$\_$(1,k2)+g$\_$(1,p2))*g$\_$(1,lor6))/(sqrt$\_$(2)*p2(k2)*(2*k2(k1)+2*p2(k1)\\
\hspace*{1.cm}	+2*p2(k2))))*(m*gi$\_$(1)+g$\_$(1,p2))*((-k1(lor1)-k2(lor1)+p1(lor1)-p2(lor1))*(-k1(lor2)\\
\hspace*{1.cm}	-k2(lor2)+p1(lor2)-p2(lor2))-d$\_$(lor1,lor2)*(m${}^{\wedge}$2+M${}^{\wedge}$2+2*k2(k1)\\
\hspace*{1.cm}	-2*p1(k1)-2*p1(k2)+2*p2(k1)+2*p2(k2)-2*p2(p1))))/(3*pi$\_$));\\
trace4,1;\\
contract 0;\\
.sort;\\
id k1.k1 = 0;\\
id k2.k2 = 0;\\
id p1.p1 = M${}^{\wedge}$2;\\
id p2.p2 = m${}^{\wedge}$2;\\
.sort;\\
$\#$call put({``}$\%$E{''}, resFL)\\
$\#$fromexternal}}

\noindent\rule[.4\baselineskip]{\textwidth}{0.1pt}
\end{samepage}
\vspace{.03cm}\noindent\includegraphics{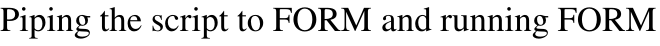}

\vspace{.03cm}\noindent\includegraphics{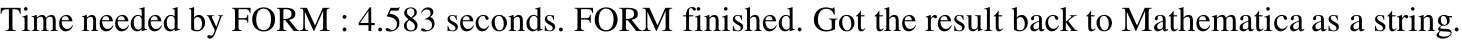}

\vspace{.03cm}\noindent\includegraphics{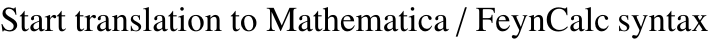}

\vspace{.03cm}\noindent\includegraphics{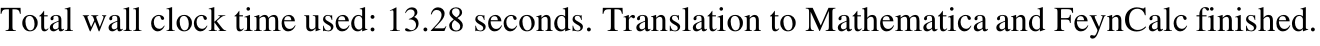}

\vspace{.03cm}\noindent\includegraphics{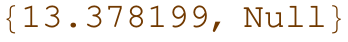}

\vspace{.03cm}\noindent\includegraphics{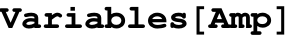}

\vspace{.03cm}\noindent\includegraphics{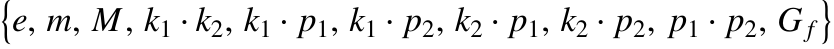}

\vspace{.03cm}\noindent\includegraphics{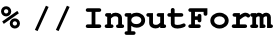}

\vspace{.03cm}\noindent\includegraphics{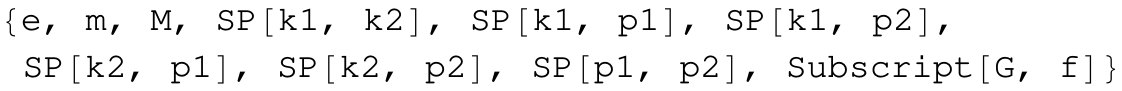}

\vspace{.03cm}\noindent\includegraphics{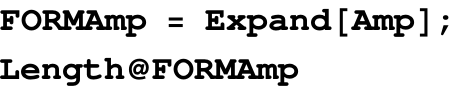}

\vspace{.03cm}\noindent\includegraphics{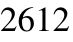}

\subsec{Check with FeynCalc}

\vspace{.03cm}\noindent\includegraphics{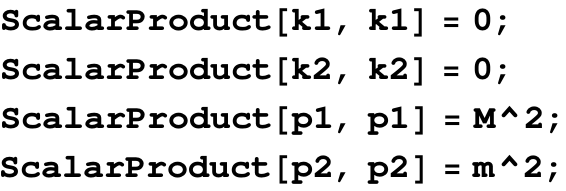}

\noindent The naive approach is too slow.

\vspace{.03cm}\noindent\includegraphics{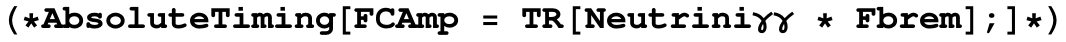}

\noindent Do this instead (also still quite slow, but good enough for checking equality)

\noindent\includegraphics{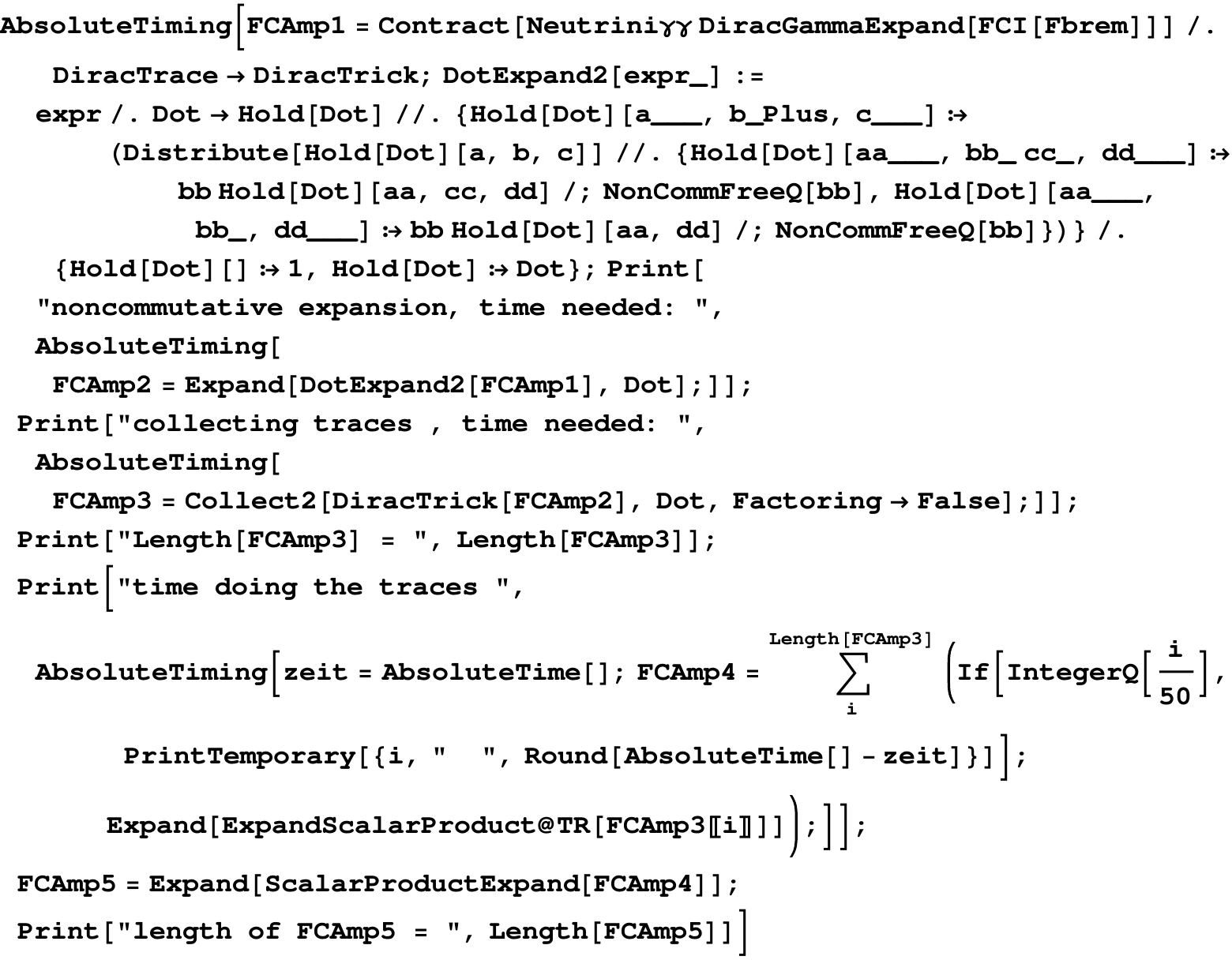}

\vspace{.03cm}\noindent\includegraphics{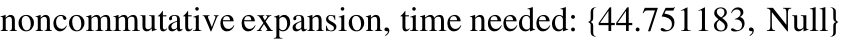}

\vspace{.03cm}\noindent\includegraphics{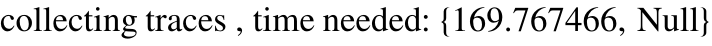}

\vspace{.03cm}\noindent\includegraphics{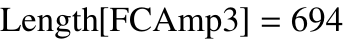}

\vspace{.03cm}\noindent\includegraphics{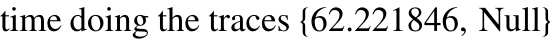}

\vspace{.03cm}\noindent\includegraphics{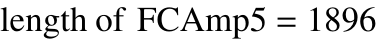}

\vspace{.03cm}\noindent\includegraphics{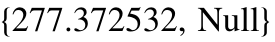}

\vspace{.03cm}\noindent\includegraphics{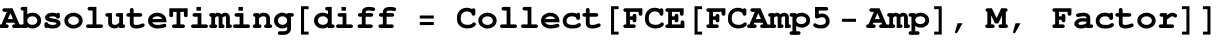}

\vspace{.03cm}\noindent\includegraphics{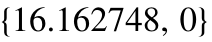}

\noindent The results of {\sc FORM} and {\sc FeynCalc} agree, but {\sc FORM} is much faster.
There is a some noticeable overhead in translating the returned string to
\textsc{FeynCalc}. In a future version of \textsc{FormLink} we will improve this
by either using parallel features of \textit{Mathematica} or by writing a
special \textit{MathLink} program similar to \texttt{FormGet.tm}.

\section{S\lowercase{ummary}}
We have implemented \textsc{FormLink} and \textsc{FeynCalcFormLink}, two
\textit{MathLink}-based programs implemented in \textit{Mathematica}, C and
\textsc{FORM}, which embed \textsc{FORM} in \textit{Mathematica} and
\textsc{FeynCalc}.
A non-\textit{MathLink}-based file-based function
\textsc{RunForm} has been also put into the \textsc{FormLink} package.
For both packages we provide configurable functions for the syntax conversions.
A limited subset of the syntaxes from both programs are automatically translated
to each other by our two \textit{Mathematica} packages.
In this way we can combine the speed of \textsc{FORM} with the generality of \textit{Mathematica}.
 A simple installation facility has been provided.

 We hope to be able to extend the functionality and range of applicability of the packages in the future.

\section{L\lowercase{icenses}}
\textsc{FormLink} and \textsc{FeynCalcFormLink} are covered by the GNU Lesser General Public License, like \textsc{FORM}.
The conditions for the use of \textsc{FORM} are laid out here: \url{http://www.nikhef.nl/~form/license/license.html} and should be
followed of course also when using \textsc{FormLink/FeynCalcFormLink}.
The license for using \textit{Mathematica} is given here:
\url{http://www.wolfram.com/legal/agreements/wolfram-mathematica.html}

\acknowledgments
We would like to thank Matteo Fael for providing us with the example about double bremsstrahlung in $\tau$ leptonic radiative decay.
Rolf Mertig would like to thank Prof Dr Thomas Gehrmann from the Institut f\"ur
Theoretische Physik, Universit\"at Z\"urich, for the invitation to teach
\textit{Mathematica} for High Energy Physics students, which encouraged him to work on \texttt{FormLink/FeynCalcFormLink}.

Feng Feng wants to thank Hai-Rong Dong, Wen-Long Sang and Prof.~Yu Jia for many useful discussions. Finally, Feng Feng would like to commemorate his beloved mother.


\begin{thebibliography}{99}
\bibitem{Kuipers:2012rf}
  J.~Kuipers, T.~Ueda, J.~A.~M.~Vermaseren and J.~Vollinga,
  FORM version 4.0,  arXiv:1203.6543 [cs.SC].  


\bibitem{Vermaseren:2000nd}
  J.~A.~M.~Vermaseren,
  math-ph/0010025.

\bibitem{Vermaseren:2011sb}
  J.~A.~M.~Vermaseren,
  FORM development,  PoS CPP {\bf 2010}, 012 (2010)  [arXiv:1101.0511 [hep-ph]].  

\bibitem{Vermaseren:2010iw}
  J.~A.~M.~Vermaseren,
  FORM facts,  Nucl.\ Phys.\ Proc.\ Suppl.\  {\bf 205-206}, 104 (2010)  [arXiv:1006.4512 [hep-ph]].  

\bibitem{Tentyukov:2008zz}
  M.~Tentyukov and J.~A.~M.~Vermaseren,
  Current status of FORM parallelization,  PoS ACAT {\bf 08} (2008) 119.  

\bibitem{Vermaseren:2008kw}
  J.~A.~M.~Vermaseren,
  The FORM project,  Nucl.\ Phys.\ Proc.\ Suppl.\  {\bf 183}, 19 (2008)  [arXiv:0806.4080 [hep-ph]].  

\bibitem{Mertig:1990an}
  R.~Mertig, M.~Bohm and A.~Denner,
  FEYNCALC: Computer algebraic calculation of Feynman amplitudes,
  Comput.\ Phys.\ Commun.\  {\bf 64} (1991) 345.

\bibitem{hep-ph/9807565}
  T.~Hahn and M.~Perez-Victoria,
  Automatized one loop calculations in four-dimensions and D-dimensions,  Comput.\ Phys.\ Commun.\ \ {\bf 118}, 153  (1999)  [hep-ph/9807565].  

\bibitem{Tentyukov:2006ys}
  M.~Tentyukov and J.~A.~M.~Vermaseren,
  Extension of the functionality of the symbolic program FORM by external software,  Comput.\ Phys.\ Commun.\  {\bf 176}, 385 (2007)  [cs/0604052 [cs-sc]].  

\bibitem{FormCourse}
Jos Vermaseren, Introduction to FORM,\\\url{http://www.nikhef.nl/~form/maindir/courses/course1/sheets5.pdf}

\bibitem{Fischer:1994pn}
  A.~Fischer, T.~Kurosu and F.~Savatier,
  QED one loop correction to radiative muon decay,
  Phys.\ Rev.\ D {\bf 49}, 3426 (1994).


\end{thebibliography}
\end{document}